\documentclass{article}

\PassOptionsToPackage{numbers, compress}{natbib}
\usepackage{paperstyle}
\renewcommand{\acksection}{\section*{Acknowledgements}}

\usepackage[utf8]{inputenc}
\usepackage[T1]{fontenc}
\usepackage{hyperref}
\usepackage{url}
\usepackage{booktabs}
\usepackage{amsfonts}
\usepackage{amsmath}
\usepackage{amssymb}
\usepackage{amsthm}
\usepackage{nicefrac}
\usepackage{microtype}
\usepackage{xcolor}
\usepackage{graphicx}
\usepackage{subcaption}
\usepackage{algorithm}
\usepackage{algpseudocode}
\usepackage{wrapfig}
\usepackage{multirow}
\usepackage{placeins}

\DeclareMathOperator*{\argmin}{arg\,min}
\DeclareMathOperator{\minSR}{minSR}
\newcommand{\Expect}{\mathbb{E}}

\newcommand{\Snorm}[1]{\left\| #1 \right\|_S^2}
\newcommand{\figureplaceholder}[2]{%
  \begin{center}
    \setlength{\fboxsep}{10pt}%
    \fbox{%
      \begin{minipage}[c][0.24\textheight][c]{0.93\linewidth}
        \centering
        \textbf{#1}\par\medskip
        \small #2
      \end{minipage}%
    }%
  \end{center}%
}

\title{Stochastic Reconfiguration as Statistical Filtering for Overparameterized Neural Quantum States}

\newcommand{\paperauthornames}{Tak Hur}
\newcommand{\paperrepository}{https://github.com/takh04/sr_filter}
\newcommand{\paperrepositorylink}{\expandafter\url\expandafter{\paperrepository}}
\author{%
  Tak Hur \\
  Department of Applied Mathematics, University of Waterloo \\
  200 University Avenue West, Waterloo, Ontario N2L 3G1, Canada \\
  \texttt{tak.hur@uwaterloo.ca}%
}

\hypersetup{
  pdftitle={Stochastic Reconfiguration as Statistical Filtering for Overparameterized Neural Quantum States},
  pdfauthor={\paperauthornames}
}
  
\begin{document}

\maketitle

\begin{abstract}
Stochastic reconfiguration (SR) is the standard optimizer for neural quantum
states (NQS), but modern NQS often have far more parameters than Monte Carlo
samples. We show that in this regime the diagonal shift is more than a numerical
stabilizer. It acts as a statistical filter for finite-sample generalization.
At a fixed wave function, SR is ridge regression from tangent features to the
centered local energy. Its residual is the \emph{expressivity gap}, the part of
imaginary-time evolution outside the current tangent space. This gap is
orthogonal to the tangent space in population, but finite batches make it act as
noise that SR can overfit. The shift therefore balances shrinkage of useful
update directions against variance from fitting sampled residuals.
Exact diagnostics on a $4\times4$ Heisenberg graph
separate two effects of overparameterization. Larger tangent spaces help when they reduce the
expressivity gap, but they can hurt when they overfit a fixed gap. In a $100$-site
transverse-field Ising family trained with a foundation NQS, validation risk is
U-shaped in the shift while variance decreases, matching the noisy-ridge model.
This view leads to multi-shift SR (MS-SR), which averages independent ridge
solves at data-adaptive shifts to form a richer, lower-variance spectral filter.
Checkpoint-local experiments show that MS-SR lowers validation risk and
update variance relative to the fixed-shift SR baseline. We further compare
MS-SR and SR in paired online training continuations, with independent
endpoint energy evaluations and a separate update-cost benchmark.
\phantomsection\label{sec:code}
The code and data for this work is available at \paperrepositorylink.
\end{abstract}

\section{Introduction}

\begin{figure}[h]
    \centering
    \IfFileExists{p_over_ns_literature.pdf}{
        \includegraphics[width=0.95\linewidth]{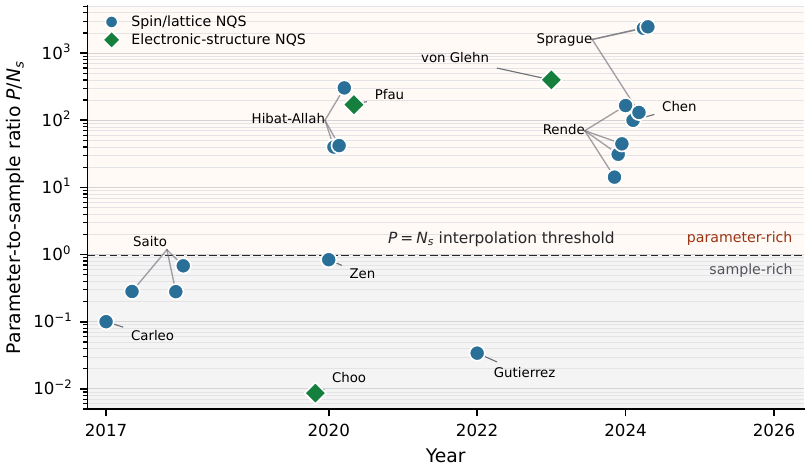}
    }{
        \figureplaceholder{Historical $P/N_s$ Regime Shift}{
            Literature scatter plot of parameter-to-sample ratios across VMC/NQS
            studies. Early RBM/NQS points are order-of-magnitude anchors; modern
            points use reported or replicated parameter and sample counts.}
    }
    \caption{Historical shift in the parameter-to-sample ratio used by VMC/NQS
    calculations. The dashed line marks \(P=N_s\). Marker shape denotes domain:
    circles are spin/lattice NQS and diamonds are electronic-structure NQS. Sources:
    \citep{carleo2017solving,saito2017bosehubbard,
    saitokato2018machine,choo2020fermionic,zen2020critical,hibatallah2020rnn,
    pfau2020ferminet,gutierrez2022real,vonglehn2023psiformer,
    rende2024simple,chen2024empowering,sprague2024variational}.}
    \label{fig:p_over_ns_literature}
\end{figure}

Stochastic Reconfiguration (SR) remains the dominant optimizer for neural quantum
states (NQS), but the regime in which it operates has changed fundamentally.
Classical variational Monte Carlo with Jastrow, Gutzwiller, and projected
mean-field ansatzes, as well as early RBM-based NQS, had modest parameter
counts \citep{mcmillan1965ground, yokoyama1987variational,
capello2005variational, tahara2008variational, sorella1998green,
sorella2001generalized, carleo2017solving}.
Kernel-form SR (minSR) removed the need to form the full $P\times P$ quantum geometric tensor (QGT), enabling
the shift to transformer-based and foundation-style wave functions with millions of parameters
\citep{choo2020fermionic, hermann2020paulinet, pfau2020ferminet,
hibatallah2020rnn, vonglehn2023psiformer, chen2024empowering,
rende2024simple, sprague2024variational, kim2024neural,
nutakki2025design, rende2025foundation}.
However, minSR still solves a dense $N_s\times N_s$ system
built from the empirical NTK matrix, at cost $O(N_s^2 P)$ to form and $O(N_s^3)$ to
solve. Increasing $N_s$ therefore remains the practical bottleneck, pushing modern
NQS from the sample-rich regime of classical VMC ($P<N_s$) into a parameter-rich regime where $P$ exceeds $N_s$ by one to three
orders of magnitude (Figure~\ref{fig:p_over_ns_literature}).

Each SR update solves a linear system with an empirical QGT, which can be
ill-conditioned when estimated from finitely many Monte Carlo
samples. The standard remedy is to add a diagonal shift before solving
\citep{sorella1998green,sorella2001generalized,carleo2017solving}. Historically, this shift was therefore understood mainly as a
numerical conditioning device. In the modern minSR regime, however, the
same operation is applied when the number of variational parameters can far exceed
the number of samples, so its role may be statistical as well as numerical.
This raises the central question of this paper: is the diagonal shift in SR
merely a numerical stabilizer, or is it a statistical filter that makes
overparameterized finite-sample updates generalize, and can this view lead to
better filters than a single $\lambda$?

This question exposes the central tension of modern NQS optimization. More expressive neural quantum states can reduce the \emph{expressivity gap}, the
component of imaginary-time evolution outside the current tangent space.
However, finite-sample SR can also overfit the part of this gap that remains.
The central contributions of this work are:

\begin{itemize}
    \item We formulate SR as a misspecified ridge-regression
    problem. The expressivity gap acts as finite-sample residual noise, yielding
    an exact held-out error identity and a spectral risk approximation for the
    bias--variance role of the diagonal shift \(\lambda\).

    \item We test this mechanism in both exactly tractable small-scale and modern
    large-scale NQS settings. Exact \(4\times4\) matched-noise and matched-state
    regression diagnostics separate finite-sample overfitting from
    expressivity-gap reduction.
    Large-scale shared-checkpoint diagnostics show the predicted U-shaped
    validation risk and monotone variance reduction as \(\lambda\) increases.

    \item We use the filter view to design multi-shift SR (MS-SR), which
    combines independent ridge solves at multiple data-adaptive shifts.
    Checkpoint-local ablations isolate the roles of multi-shift spectral
    filtering, independent-batch averaging, and learned stacking. We further
    compare MS-SR and SR in paired online training continuations, with
    independent endpoint energy evaluations and measurements of update cost.

\end{itemize}

Section~\ref{sec:ridge} develops the noisy ridge-regression view of SR.
Sections~\ref{sec:small} and~\ref{sec:large_scale} validate its predictions in
exact small-scale and modern large-scale settings. Section~\ref{sec:mssr} shows
how this perspective leads to MS-SR, with checkpoint-local ablations and
online comparisons against SR.

\section{SR as Misspecified Ridge Regression}\label{sec:ridge}

Throughout this section we fix a checkpoint \(\theta\) and assume that the wavefunction and Hamiltonian are real in the
sampling basis for notational simplicity. The complex case is obtained by replacing transposes with
Hermitian conjugates (see Appendix~\ref{app:proofs} for detail).

\paragraph{SR as tangent-space regression.}
Let $\psi_\theta$ be a neural quantum state with Born distribution
\(\pi(x)=|\psi_\theta(x)|^2/\|\psi_\theta\|^2\). Define the centered tangent
features and centered local energy by
\[
    O_c(x)=\nabla_\theta\log\psi_\theta(x)
    -\Expect_\pi[\nabla_\theta\log\psi_\theta],
    \qquad
    H_{\mathrm{loc},c}(x)=H_{\mathrm{loc}}(x)-E(\theta),
\]
where \(H_{\mathrm{loc}}(x)=\langle x|H|\psi_\theta\rangle/
\langle x|\psi_\theta\rangle\) and
\(E(\theta)=\Expect_\pi[H_{\mathrm{loc}}]\). Under normalized imaginary-time
evolution, the log amplitude changes as
\(\Delta\log\psi_\theta(x)=-\eta H_{\mathrm{loc},c}(x)+O(\eta^2)\). A parameter
update \(\theta^+=\theta-\eta\delta\) changes the log amplitude as
\(\log\psi_{\theta^+}(x)-\log\psi_\theta(x)
=-\eta O_c(x)^T\delta+O(\eta^2)\), up to an \(x\)-independent normalization
term. Matching these first-order changes shows that fixed-checkpoint SR fits
tangent predictions \(O_c(x)^T\delta\) to the centered local energy \(H_{\mathrm{loc},c}(x)\).

Writing \(S^\dagger\) for the Moore--Penrose pseudoinverse of
\(S=\Expect_\pi[O_c(x)O_c(x)^T]\), and setting
\(g=\Expect_\pi[O_c(x)H_{\mathrm{loc},c}(x)]\), the ideal fixed-checkpoint SR
direction is the population least-squares solution
\begin{equation}
\label{eq:population_sr_regression}
    \delta^*
    =
    S^\dagger g
    \in
    \argmin_\delta
    \Expect_\pi\!\left[
        \bigl(O_c(x)^T\delta-H_{\mathrm{loc},c}(x)\bigr)^2
    \right].
\end{equation}

\paragraph{Empirical SR as misspecified ridge regression.}
The centered local energy need not be exactly representable by the current
tangent features. We define the tangent-space expressivity gap
$
    \epsilon(x)
    =
    H_{\mathrm{loc},c}(x)-O_c(x)^T\delta^* .
$
The normal equations imply \(\Expect_\pi[\epsilon(x)O_c(x)]=0\), and the gap
variance is
$
    \sigma_{\mathrm{gap}}^2
    =
    \Expect_\pi[\epsilon(x)^2]
    =
    \operatorname{Var}_\pi(H_{\mathrm{loc}})
    -
    g^T S^\dagger g .
$
We call \(\epsilon(x)\) the misspecification noise: it is the part of the
imaginary-time target that lies outside the current tangent space.

Given a Monte Carlo batch \(\mathcal D=\{x_j\}_{j=1}^{N_s}\), empirical SR solves
the ridge problem
\[
    \hat\delta_\lambda=(\hat S+\lambda I)^{-1}\hat g,
    \qquad
    \hat S=\frac1{N_s}\sum_{j=1}^{N_s}O_c(x_j)O_c(x_j)^T,
    \qquad
    \hat g=\frac1{N_s}\sum_{j=1}^{N_s}O_c(x_j)H_{\mathrm{loc},c}(x_j).
\]
Although \(\epsilon(x)\) is orthogonal to the tangent features in population, it
is still present in the sampled targets \(H_{\mathrm{loc},c}(x_j)\). An
overparameterized empirical solve can therefore memorize this residual on the
batch, even though it does not correspond to a generalizable tangent-space
update. In this sense, the expressivity gap behaves as regression noise.

\paragraph{Excess risk analysis.}
Here, we are interested in \(S\)-norm excess risk
$
    \mathcal E_\lambda
    =
    \Expect_{\mathcal D}
    \|\hat\delta_\lambda-\delta^*\|_S^2,
$
where the expectation is over the Monte Carlo batch used to form
\(\hat\delta_\lambda\). This is the out-of-sample error of the checkpoint-local
ridge problem with features \(O_c(x)\), target \(H_{\mathrm{loc},c}(x)\), and
residual noise \(\epsilon(x)\).

The exact bias--variance identity, together with the spectral proxy (see Appendix~\ref{app:proofs} for details), gives
\begin{equation}\label{eq:spectral_risk}
    \mathcal E_\lambda =
    \underbrace{
    \Snorm{\Expect_{\mathcal D}[\hat\delta_\lambda]-\delta^*}
    }_{\text{Bias}}
    +
    \underbrace{
    \Expect_{\mathcal D}
    \Snorm{\hat\delta_\lambda-\Expect_{\mathcal D}[\hat\delta_\lambda]}
    }_{\text{Variance}}
    \approx
    \underbrace{
    \sum_i
    s_i
    \left(\frac{\lambda}{s_i+\lambda}\right)^2
    (\beta_i^*)^2
    }_{\mathcal B_{\mathrm{shrink}}(\lambda)}
    +
    \underbrace{
    \frac{\sigma_{\mathrm{gap}}^2}{N_s}
    \sum_i
    \left(\frac{s_i}{s_i+\lambda}\right)^2
    }_{\mathcal V_{\mathrm{gap}}(\lambda)} ,
\end{equation}
where \(S=V\operatorname{diag}(s_i)V^T\) and
\(\beta_i^*=(V^T\delta^*)_i\). The approximation replaces the empirical
inverse by its population counterpart and uses a scalar residual-noise
covariance. The first term is shrinkage bias: it grows when
\(\lambda\) damps useful components of \(\delta^*\). The second term is
finite-sample variance: it decreases because the same filter suppresses fits to
sampled gap residuals. Thus \(\lambda\) is a statistical regularizer, not only a
numerical tolerance. Standard SR, however, is restricted to the one-parameter
filter family \(s/(s+\lambda)\), motivating the multi-\(\lambda\) filters in
Section~\ref{sec:mssr}.

This \(S\)-norm excess risk is physically relevant because it is closely related to the infidelity, a standard distance measure between quantum states. For bounded updates,
\begin{equation}\label{eq:fs_local_metric}
    \mathcal I_\theta(\delta,\delta')
    =
    1-
    \left|
    \left\langle
        \psi_{\theta-\eta\delta}
        \middle|
        \psi_{\theta-\eta\delta'}
    \right\rangle
    \right|^2
    =
    \eta^2\Snorm{\delta-\delta'}
    +
    O(\eta^3).
\end{equation}
Therefore \(\mathcal E_\lambda\) controls the expected local infidelity between
the ideal population update and the empirical SR update, up to the factor
\(\eta^2\) and higher-order terms.

\paragraph{Large-system diagnostics for excess risk and variance.}
For small systems, the population quantities \(S\), \(\delta^*\),
\(\sigma_{\mathrm{gap}}^2\), and
\(\|\hat\delta_\lambda-\delta^*\|_S^2\) can be computed by exact diagonalization.
This is impossible in the large scale regimes where modern NQS are tuned
and compared. We therefore use two fixed-checkpoint diagnostics that require
only held-out predictions.

The first diagnostic is a validation residual. Let
\(\mathcal D_{\mathrm{val}}\) be independent of the batches used to compute the
SR updates. Since
\(H_{\mathrm{loc},c}(x)=O_c(x)^T\delta^*+\epsilon(x)\) with
\(\Expect_\pi[\epsilon(x)O_c(x)]=0\), any fixed update \(\delta\) satisfies
\begin{equation}\label{eq:rval_decomp}
    R_{\mathrm{val}}(\delta)
    =
    \frac1{|\mathcal D_{\mathrm{val}}|}
    \sum_{x\in\mathcal D_{\mathrm{val}}}
    \bigl(O_c(x)^T\delta-H_{\mathrm{loc},c}(x)\bigr)^2,
    \qquad
    \Expect_{\mathcal D_{\mathrm{val}}}
    \left[R_{\mathrm{val}}(\delta)\right]
    =
    \|\delta-\delta^*\|_S^2+\sigma_{\mathrm{gap}}^2 .
\end{equation}
Thus, for independent SR updates
\(\hat\delta_{\lambda,1},\dots,\hat\delta_{\lambda,m}\) at the same checkpoint,
we use
\(\widehat{\mathcal E}_{\mathrm{val}}(\lambda)
=m^{-1}\sum_{j=1}^m R_{\mathrm{val}}(\hat\delta_{\lambda,j})\). Its expectation
is \(\mathcal E_\lambda+\sigma_{\mathrm{gap}}^2\), so it tracks excess risk plus the expressivity-gap.

The second diagnostic measures how much the update varies across independent
training batches. With
\(\bar\delta_\lambda=m^{-1}\sum_j\hat\delta_{\lambda,j}\), define
\begin{equation}\label{eq:vmb_def}
    \mathcal V_{\mathrm{mb}}(\lambda)
    =
    \frac1{m-1}
    \sum_{j=1}^m
    \frac1{|\mathcal D_{\mathrm{val}}|}
    \sum_{x\in \mathcal D_{\mathrm{val}}}
    \bigl(O_c(x)^T(\hat\delta_{\lambda,j}-\bar\delta_\lambda)\bigr)^2 .
\end{equation}
As \(m\) and the validation batch grow, this converges to the variance component
of the excess risk. Both diagnostics are practical at large scale because they only require
held-out tangent-feature predictions. Since they do not require forming or
solving a validation QGT or NTK system, they can be evaluated on validation
batches much larger than the SR training batch with modest computational cost.

\section{Small-scale Dissection through Regression}\label{sec:small}

\begin{figure}[t]
    \centering
    \includegraphics[width=0.95\textwidth]{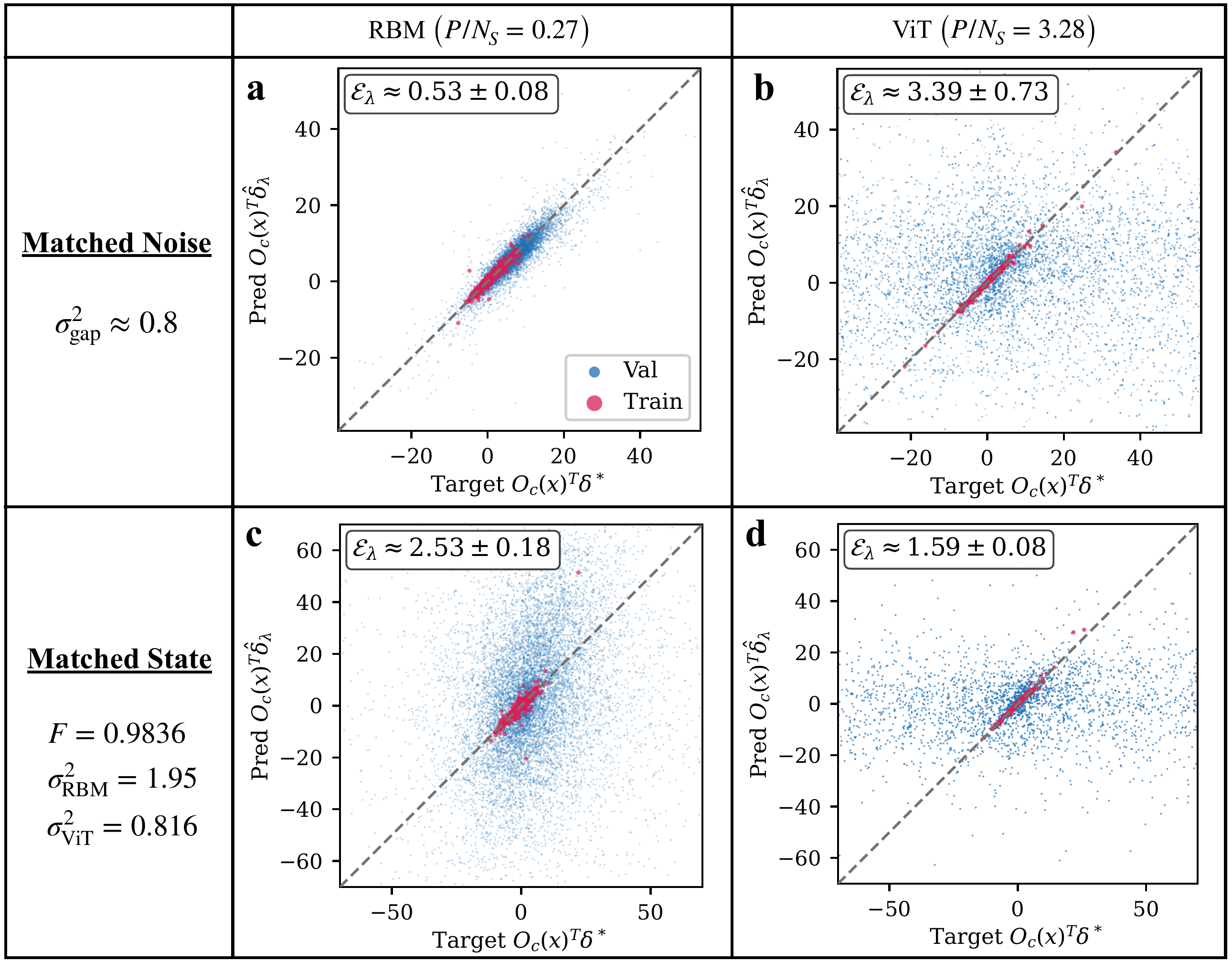}
    \caption{Exact regression diagnostics on the \(4\times4\)
    Heisenberg graph specified in Appendix~\ref{app:small}, at \(N_s=4096\)
    and \(\lambda=10^{-9}\). Panels compare empirical and population-optimal
    amplitude tangent predictions. Gap variance and excess risk refer to
    this regression. Matched noise exposes ViT
    overfitting at comparable residual variance; matched state shows lower
    gap variance and excess risk for the ViT.}
    \label{fig:smallscale_paradox}
\end{figure}

This section tests the regression mechanism of Section~\ref{sec:ridge}
through the real-amplitude component of complex neural quantum states.
The experiment uses a $4\times4$ spin-$1/2$ Heisenberg graph,
\[
 H=J_1\sum_{(i,j)\in\mathcal B_1}\mathbf S_i\cdot\mathbf S_j
   +J_2\sum_{(i,j)\in\mathcal B_2}\mathbf S_i\cdot\mathbf S_j,
 \qquad J_1=1,\quad J_2=0.5,
\]
where $\mathbf S_i=\boldsymbol\sigma_i/2$. Gap and risk values below concern the real-amplitude
channel, where phase contributions are excluded.

We compare an RBM with $P/N_s=0.27$ and a ViT with $P/N_s=3.28$, using
$N_s=4096$ and the near-ridgeless shift $\lambda=10^{-9}$ to expose fitting
of sampled residuals. The matched-noise and matched-state comparisons use
$100$ independent Born-resampled batches.
Exact summation over all $2^{16}$ states gives the population reference and
gap variance. We report the mean and standard error of the
regression excess risk.
Figure~\ref{fig:smallscale_paradox} visualizes the corresponding empirical
and population-optimal amplitude tangent predictions.

The experiment separates two effects that are coupled.
The matched-noise comparison keeps $\sigma_{\mathrm{gap}}^2$ approximately fixed
and tests the finite-sample cost of a larger tangent space. The matched-state
comparison keeps the represented wavefunction approximately fixed and tests the
expressivity benefit of the larger tangent space. Full model and optimization
details are given in Appendix~\ref{app:small}.

\paragraph{Matched-noise comparison.}
To test the finite-sample cost of overparameterization, we select independently
trained RBM and ViT checkpoints with comparable expressivity-gap variance. In
Figure~\ref{fig:smallscale_paradox} panel (a) and (b), both values of
$\sigma_{\mathrm{gap}}^2$ are close to $0.8$. At
$\lambda=10^{-9}$, the RBM remains close to the population-optimal amplitude direction, with
$\mathcal E_\lambda\approx0.53\pm0.08$. The ViT has comparable
misspecification noise, but its empirical solve fits the sampled residual much
more strongly, increasing the excess-risk estimate to
$\mathcal E_\lambda\approx3.39\pm0.73$. This illustrates the regression-overfitting mechanism
underlying Eq.~\eqref{eq:spectral_risk}. At comparable amplitude gap variance,
the larger tangent space gives the empirical regression more directions in
which to fit sample-specific noise.

\paragraph{Matched-state comparison.}
To isolate the benefit of a larger tangent space, we freeze a partially
converged RBM checkpoint and fit a ViT to represent the same quantum state by
minimizing infidelity~\cite{sinibaldi2023unbiasing,gravina2410neural}. The two represented states have fidelity \(F\geq0.98\).
Thus the wavefunction is nearly fixed, while the tangent space changes.
Under this matched-state protocol, ViT's larger tangent space lowers
the amplitude gap variance from \(\sigma_{\mathrm{gap}}^2=1.95\) to \(0.816\).
The excess-risk estimate also decreases from
\(\mathcal E_\lambda\approx2.53\pm0.19\) to
\(\mathcal E_\lambda\approx1.59\pm0.08\). Thus overparameterization can improve
regression generalization when its additional tangent directions reduce enough
misspecification noise.

Together, these comparisons illustrate the corresponding tension in regression: a larger tangent space can help when gap reduction outweighs extra
finite-sample error, but can hurt when the solve overfits a comparable gap.
This motivates the statistical-filtering interpretation of the diagonal shift
$\lambda$.

\section{Large-Scale Checkpoint Diagnostics of the Noisy-Ridge Mechanism}
\label{sec:large_scale}

\begin{figure}[t]
    \centering
    \includegraphics[width=0.95\textwidth]{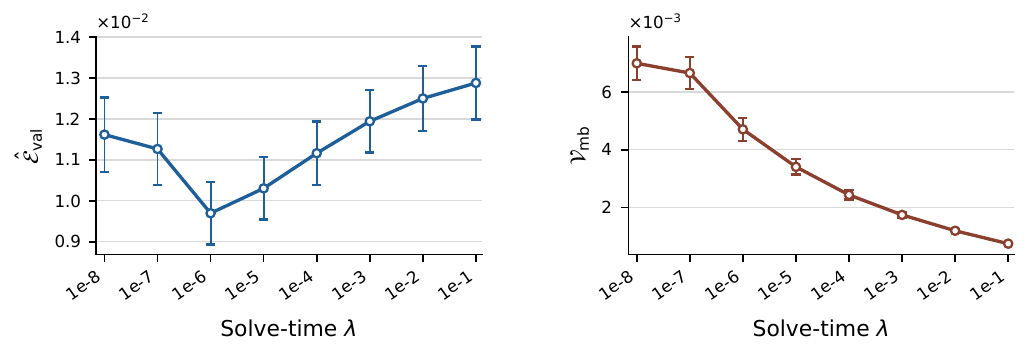}
    \caption{Large-scale shared-checkpoint test of the noisy-ridge mechanism on
    the periodic \(L=100\) transverse-field Ising model with \(h\) ranging from \(0.8\) to \(1.2\) with foundational NQS.
    A fixed set of late-training checkpoints is reused while only the
    solve-time shift $\lambda$ is varied. $\widehat{\mathcal E}_{\mathrm{val}}$ is U-shaped, while $\mathcal V_{\mathrm{mb}}$ decreases monotonically with
    $\lambda$, matching the two predictions of Section~\ref{sec:ridge}.}
    \label{fig:large_scale_ucurve}
\end{figure}

The exact $4\times4$ study isolates the noisy-ridge mechanism using population
quantities. We next test whether the same mechanism remains visible in modern
overparameterized NQS, where the population QGT, population SR direction, and
expressivity-gap variance are unavailable. We therefore test the mechanism using the
checkpoint-local diagnostics introduced in Section~\ref{sec:ridge}.

We use a foundation neural quantum state (FNQS) on the $L=100$ periodic
one-dimensional transverse-field Ising model (TFIM) family,
$
    H=-h\sum_i X_i-\sum_i Z_iZ_{i+1},
$
with $h\in[0.8,1.2]$. A single conditional
wave function $\log\psi_\theta(x;h)$ is trained across many Hamiltonian
parameters, so the SR update must aggregate information over a family of
Hamiltonians. The model has
$P=198{,}144$ parameters, while each SR step uses a batch of
$N_s=12{,}000$ samples, split over $6000$ training $h$ values with two spin
configurations per field. We use the same architectural and sampler setting as
\citet{rende2025foundation} (see Appendix~\ref{app:large} for experimental details).

The diagnostic experiment uses a shared-checkpoint protocol. We first take ten
late-training checkpoints from the run trained with
$\lambda_{\mathrm{train}}=10^{-4}$. These checkpoints are then held fixed. At
each checkpoint, we perform an offline sweep over the solve-time shift
$
    \lambda\in\{10^{-8},10^{-7},\ldots,10^{-1}\},
$
recomputing standard SR updates on fresh resampled batches for each value of
$\lambda$. This design keeps the wave function, tangent geometry, and
expressivity-gap fixed while varying only the ridge penalty. It therefore
tests the local bias--variance mechanism directly, not the different training trajectories.

For each checkpoint and each diagnostic shift, we construct $m=100$ independent
standard-SR updates from independently resampled training batches. The validation
diagnostic uses $120{,}000$ held-out samples, consisting of $6000$ held-out
$h$ values, disjoint from the training $h$ values, and $20$ spin configurations per
$h$ value. Predictions are centered separately for each $h$ value, since each
Hamiltonian parameter has its own local-energy distribution. The validation pass
then estimates both $\widehat{\mathcal E}_{\mathrm{val}}$ and
$\mathcal V_{\mathrm{mb}}$ without forming or solving any validation QGT system.

Figure~\ref{fig:large_scale_ucurve} shows the fixed-checkpoint signature
predicted by the noisy-ridge view. The
$\hat{\mathcal{E}}_{\mathrm{val}}$ curve is U-shaped in \(\lambda\), while
\(\mathcal V_{\mathrm{mb}}\) decreases as \(\lambda\) increases. Since all
shifts are evaluated on the same checkpoints, this trend reflects the
solve-time ridge filter rather than different training trajectories. Small
shifts leave useful directions nearly undamped but fit more batch-dependent
noise. Large shifts reduce this variance but also shrink useful population
directions.

\section{Multi-Shift SR for Statistical Filtering and Variance Reduction}
\label{sec:mssr}

Motivated by the noisy-ridge view, we introduce multi-shift SR (MS-SR), an
optimizer that replaces the single shifted SR solve with a mixture of shifted
SRs computed on independent batches. Standard SR uses one ridge scale
and therefore applies the same spectral filter \(s/(s+\lambda)\) to every QGT
eigendirection. MS-SR instead combines several regularization scales, while the
independent batches decorrelate finite-sample errors.

At checkpoint \(\theta_t\), MS-SR draws \(K\) independent batches
\(\mathcal D_1,\ldots,\mathcal D_K\), chooses shifts \(\lambda_k\) from
empirical NTK-spectrum quantiles, and computes candidate updates
\(\hat\delta_k=\minSR(\mathcal D_k,\lambda_k)\). It then forms
\(\hat\delta_{\mathrm{MS}}=\sum_{k=1}^K w_k\hat\delta_k\), with
\(\mathbf w\in\Delta^{K-1}\). The weights are selected by leave-one-batch-out
stacking, so each candidate update is evaluated only on batches that were not
used for its ridge solve. The full procedure is given in
Algorithm~\ref{alg:mssr_appendix} in Appendix~\ref{app:mssr}.

This construction has a simple bias--variance interpretation. For fixed shifts
and mixture weights, define
$
    \bar\delta_k
    =
    \Expect_{\mathcal D_k}
    \left[
        \hat\delta_k
    \right].
$
Then the excess risk decomposes exactly as
\begin{equation}
\label{eq:mssr_decomp}
    \Expect
    \left[
        \left\|
            \hat\delta_{\mathrm{MS}}-\delta^*
        \right\|_S^2
    \right] =
    \left\|
        \sum_{k=1}^K
        w_k\bar\delta_k
        -
        \delta^*
    \right\|_S^2 +
    \Expect
    \left[
        \left\|
            \sum_{k=1}^K
            w_k
            \left(
                \hat\delta_k
                -
                \bar\delta_k
            \right)
        \right\|_S^2
    \right].
\end{equation}
The first term is the bias of the average filter. The second term is the
finite-batch variance of the average update. MS-SR targets both terms: multiple
shifts enlarge the filter class, while independent batches reduce the variance
of the noisy solves.

\paragraph{Why multiple shifts help.}
\label{par:multi_lambda}

Consider the population spectral approximation from
Section~\ref{sec:ridge}. If
\(S=V\operatorname{diag}(s_i)V^T\), then a single ridge solve applies the
filter
$
    f_{\lambda}(s)
    =
    {s}/{(s+\lambda)}
$
to each eigendirection. A convex mixture of ridge solves applies instead
$
    f_{\mathbf w,\boldsymbol\lambda}(s)
    =
    \sum_k
    w_k
    {s}/{(s+\lambda_k)}.
$
Thus MS-SR is still a spectral filter, but it is no longer restricted to the
single-shift family.

The difference is clearest at the two ends of the spectrum. For small \(s\),
\(f_{\mathbf w,\boldsymbol\lambda}(s)
=s\sum_k w_k/\lambda_k+O(s^2)\). For large \(s\),
\(f_{\mathbf w,\boldsymbol\lambda}(s)
=1-(\sum_k w_k\lambda_k)/s+O(s^{-2})\).
With a single shift, the same \(\lambda\) controls both the suppression of
low-eigenvalue directions and the residual shrinkage of high-eigenvalue
directions. On the other hand, these two behaviors can be tuned separately with multiple shifts. 
This can improve the aggregate bias--variance tradeoff when useful update directions
and noise-sensitive directions are not controlled well by the same shift.
\paragraph{Why independent batches help.}
\label{par:multi_batch}

The second term in Eq.~\eqref{eq:mssr_decomp} depends on the correlations
between candidate updates. If the candidates are computed on independent
batches, the cross terms vanish and
$
    \Expect
        \|
            \sum_k
            w_k
            (
                \hat\delta_{\lambda_k}^{(k)}
                -
                \bar\delta_{\lambda_k}
            )
        \|_S^2
    =
    \sum_k
    w_k^2
    \hspace{0.1em}
    \Expect
        \|
            \hat\delta_{\lambda_k}^{(k)}
            -
            \bar\delta_{\lambda_k}
        \|_S^2.
$
For uniform weights, this is a factor-\(K\) reduction of the average
single-batch variance in the ideal independent-batch limit. Within-batch autocorrelation and residual cross-batch
covariance are discussed in Appendix~\ref{app:sampling}.

Independent batches produce a bagging-like effect by decorrelating the sampling
noise across candidate updates. If all \(K\) shifts are solved on the same
batch, the resulting candidates can still enrich the spectral filter, but their
sampling errors remain shared, so the independent-batch factor-$K$ reduction
no longer applies.
A single larger batch of size \(KN_s\) would also reduce variance, but it is substantially more expensive
computationally, with dense solve cost \(O(K^3N_s^3)\). In contrast,
\(K\) independent solves at batch size \(N_s\) cost only \(O(KN_s^3)\) and can
be run in parallel. This targets variance reduction without the cubic solve
cost of one enlarged batch. At $K=4$, the single-GPU TFIM benchmark gives
MS-SR approximately four times the runtime of a single SR update and about
$1\%$ overhead over bagged SR (Figure~\ref{fig:cost_and_online}(c);
Appendix~\ref{app:cost_protocol}).

\paragraph{Shift and weight selection.}
\label{par:mssr_weights}

MS-SR first chooses the shifts and then fits the mixture weights. In our
\(K=4\) experiments, the shifts are the empirical NTK-spectrum quantiles
\(\{0.9,0.7,0.4,0.1\}\). These quantiles place the ridge transition scale at several parts of the
empirical spectrum, producing candidate updates with different bias--variance
tradeoffs rather than relying on a single regularization.

Given these shifts, Algorithm~\ref{alg:mssr_appendix} uses leave-one-batch-out stacking
to choose the weights. Each candidate update is evaluated on batches that were
not used for its ridge solve. Minimizing the residual between tangent-feature
predictions and centered local energies gives a data-dependent cross-validation
criterion.

\begin{figure}[!ht]
    \centering
    \includegraphics[width=\textwidth]{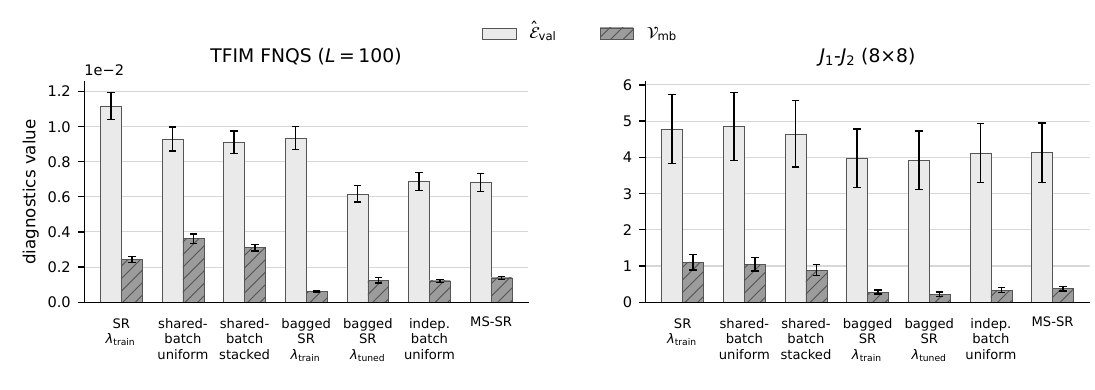}
    \caption{Checkpoint-local ablations for the ingredients of MS-SR at $K=4$.
    Paired bars show raw $\hat{\mathcal{E}}_\mathrm{val}$ and $\mathcal{V}_{\mathrm{mb}}$
    averaged over fixed source checkpoints with one standard error. Standard SR
    uses $\lambda_{\mathrm{train}}=10^{-4}$. Both systems show bagged SR at
    $\lambda_{\mathrm{train}}$ and at $\lambda_{\mathrm{tuned}}$, selected per
    checkpoint by an eight-point bagged-SR residual sweep using the evaluation data.
    All multi-shift variants use the same NTK quantile grid.}
    \label{fig:mssr_diagnostics}
\end{figure}
\FloatBarrier

\paragraph{Ablation study of MS-SR.}
\label{par:mssr_ablations}

Figure~\ref{fig:mssr_diagnostics} examines three ingredients: independent-batch
averaging, shift diversity, and learned weights. We compare validation residuals
and update variance at fixed TFIM FNQS and $8\times8$ $J_1$-$J_2$ ViT
checkpoints ($J_2/J_1=0.5$). Standard SR provides the single-update reference;
each ensemble combines four candidates:

\begin{center}
\small
\setlength{\tabcolsep}{4pt}
\begin{tabular}{@{}llll@{}}
\toprule
Method & Candidate fitting data & Shifts & Weights \\
\midrule
Standard SR & One batch & Training shift & Single update \\
Bagged SR & Four independent batches & One common shift & Uniform \\
Shared-batch multi-shift SR & One shared batch & Four shifts & Uniform \\
Shared-batch stacked SR & One shared batch & Four shifts & Learned \\
Independent-batch multi-shift SR & Four independent batches & Four shifts & Uniform \\
MS-SR & Four independent batches & Four shifts & Learned \\
\bottomrule
\end{tabular}
\end{center}

Bagging tests independent averaging at one shift. Shared-batch variants mix
shifts with shared sampling noise. Uniform and learned weights use identical
candidates within each construction to test stacking. MS-SR combines all three
ingredients. All multi-shift variants use the same NTK-quantile grid. Stacking
fits weights on samples held out from the corresponding candidate solves.

Figure~\ref{fig:mssr_diagnostics} shows bagging at $\lambda_{\mathrm{train}}$
and a checkpoint-specific $\lambda_{\mathrm{tuned}}$. The latter minimizes the bagged residual on the
reporting data, making it an oracle reference. 
Appendix~\ref{app:mssr} gives the fitting and tuning protocols.

Relative to standard SR at the training shift, MS-SR reduces
$\widehat{\mathcal E}_{\mathrm{val}}$ and $\mathcal V_{\mathrm{mb}}$ by
$39\%$/$43\%$ on TFIM and $14\%$/$66\%$ on $J_1$-$J_2$.
Bagging at $\lambda_{\mathrm{tuned}}$ reduces them by $45\%$/$49\%$ and
$18\%$/$80\%$, respectively, yielding lower mean values than MS-SR in both
systems. Shared-batch multi-shift SR reduces the TFIM residual by $17\%$,
but increases the $J_1$-$J_2$ residual by $2\%$.

The TFIM comparison also illustrates the role of shift selection. Relative to
bagging at $\lambda_{\mathrm{train}}$, MS-SR has $27\%$ lower residual, with
the same ordering on all ten checkpoints. Bagging at that larger shift has
lower variance but higher total residual, illustrating the cost of shrinkage bias.
The fixed-shift sweep in Appendix~\ref{app:robustness} shows that bagging
achieves a lower residual than MS-SR near $10^{-6}$--$10^{-7}$. Thus a
well-chosen single shift can suffice, while MS-SR offers adaptation without
a validation-based shift search.

On $J_1$-$J_2$, bagging at $\lambda_{\mathrm{train}}$ has mean residual
$3.98$, compared with $3.92$ at $\lambda_{\mathrm{tuned}}$ and $4.13$ for
MS-SR. Tuning reduces the mean residual by a further $1.5\%$
relative to training-shift bagging, selecting shifts from $10^{-4}$ to $10^{-1}$.

\paragraph{Robustness and choice of $K$.}
Appendix~\ref{app:robustness} reports $K\in\{2,3,4,5\}$, two alternative
quantile grids, and checkpoints trained at $10^{-3}$ and $10^{-5}$.
MS-SR's TFIM residual decreases from $0.00854$ to $0.00638$ as $K$ grows,
with diminishing gains and one extra batch and solve per candidate. At every
$K$ it beats bagging at the training shift, while bagging at
$\lambda_{\mathrm{tuned}}$ remains better. The alternative grids change the residual by at most
$3.4\%$, and alternative training shifts give near-parity with that bagged baseline.

\paragraph{Online training and computational cost.}
\label{par:mssr_equal_budget}
With $K=4$, MS-SR combines four candidate SR updates at each optimization
step. Figure~\ref{fig:cost_and_online}(a) therefore reports progress in
nominal SR-equivalent updates: one standard-SR step counts as one unit and
one MS-SR step as four. Under this convention, 125 MS-SR steps and 500 SR
steps both reach 500 units. This axis counts candidate updates;
implementation-dependent computational costs are assessed separately
(Appendices~\ref{app:historical_online} and~\ref{app:cost_protocol}).

We examine online behavior through a post hoc comparison of five paired
continuations from a single shared SR checkpoint at iteration $2500$ on the
$8\times8$ $J_1$--$J_2$ model, using a ViT ansatz without explicit
lattice-symmetry symmetrization. The small expressivity gap early in training
motivates using standard SR initially and reserving MS-SR's additional cost
for later refinement. Both methods use the same batch size and learning rate.
Full model, training, and evaluation details are given in Appendix~\ref{app:historical_online}.

Independent endpoint evaluations give MS-SR lower energy in
3 of 5 pairs (Figure~\ref{fig:cost_and_online}(b)), with mean
$\Delta E/N=-1.78\times10^{-5}$ and paired $95\%$ Student-$t$ interval
$[-7.05,\,3.50]\times10^{-5}$. The interval includes zero and summarizes
variation across paired differences conditional on this source state.
Additional online update-tracking diagnostics against a full-summation
population SR reference are reported in Appendix~\ref{app:online_tracking}.

\begin{figure}[!htb]
    \begin{minipage}[t]{0.48\textwidth}
        \vspace{0pt}\centering
        {\small\textbf{(a) $J_1$--$J_2$ training loss history}}\par\smallskip
        \includegraphics[width=\linewidth]{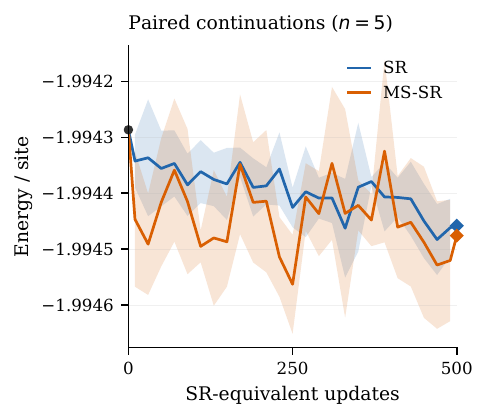}
    \end{minipage}\hfill
    \begin{minipage}[t]{0.49\textwidth}
        \vspace{0pt}\centering
        {\small\textbf{(b) $J_1$--$J_2$ endpoint energies/site}}\par\smallskip
        {\setlength{\tabcolsep}{3pt}
\footnotesize
\begin{tabular}{@{}rrrr@{}}
\toprule
Pair & SR & MS-SR & $\Delta\;(10^{-5})$ \\
\midrule
1 & $\mathbf{-1.9944675}$ & $-1.9944307$ & $+3.69 \pm 1.09$ \\
2 & $-1.9944450$ & $\mathbf{-1.9945160}$ & $-7.10 \pm 1.07$ \\
3 & $\mathbf{-1.9944733}$ & $-1.9944722$ & $+0.11 \pm 1.11$ \\
4 & $-1.9944593$ & $\mathbf{-1.9944668}$ & $-0.75 \pm 1.04$ \\
5 & $-1.9944451$ & $\mathbf{-1.9944934}$ & $-4.83 \pm 1.02$ \\
\midrule
Mean & $-1.9944580$ & $-1.9944758$ & $-1.78$ \\
\bottomrule
\end{tabular}
}\par\medskip
        {\small\textbf{(c) TFIM update-construction cost}}\par\smallskip
        {\footnotesize\setlength{\tabcolsep}{4pt}
        \begin{tabular}{@{}lrrr@{}}
            \toprule
            Method & \shortstack{Batches/\\solves} & \shortstack{Time\\(s)} & \shortstack{vs.\\SR} \\
            \midrule
            SR & 1/1 & 13.38 & $1.00\times$ \\
            Bagged SR & 4/4 & 53.55 & $4.00\times$ \\
            MS-SR & 4/4 & 54.09 & $4.04\times$ \\
            \bottomrule
        \end{tabular}}
    \end{minipage}
    \caption{Online performance results for MS-SR against SR.
    \textbf{(a)} Five paired continuations from a single shared SR checkpoint.
    Curves show means of binned traces with $\pm1$ across-run sample
    standard deviation. The nominal horizontal axis counts candidate SR
    updates: one per SR step and four per MS-SR step. Diamonds at 500 mark
    independent endpoint means.
    \textbf{(b)} Endpoint energies for the same five paired
    continuations as in (a). Bold marks the lower energy.
    $\Delta=(E_{\mathrm{MS\text{-}SR}}-E_{\mathrm{SR}})/N$. Individual
    $\pm$ values are combined Monte Carlo standard errors in units of
    $10^{-5}$.
    \textbf{(c)} A separate TFIM benchmark on one H100 80GB GPU at $K=4$,
    showing medians over three paired device/seed settings for the
    NTK-quantile MS-SR implementation. Protocols and computational accounting
    are given in Appendices~\ref{app:cost_protocol}
    and~\ref{app:historical_online}.}
    \label{fig:cost_and_online}
\end{figure}
\FloatBarrier

\section{Related Work}

Neural quantum states have evolved rapidly from compact RBM-like ans\"atze to
deep convolutional, recurrent, transformer-based, fermionic, and
foundation-style wave functions
\citep{carleo2017solving, hibatallah2020rnn, pfau2020ferminet,
vonglehn2023psiformer, nomura2021dirac, rende2024simple,
nutakki2025design, rende2025foundation}. Autoregressive NQS provide a
complementary route to scalability because their normalized factorization
permits sampling from the model Born distribution rather than
Markov-chain sampling \citep{hibatallah2020rnn}. This changes the sampling and
autocorrelation structure of VMC, but it does not remove the finite-sample
nature of SR, since QGTs and forces are still estimated from finite samples.

SR is the natural-gradient method of VMC, with the QGT playing the role of the
Fisher metric on the variational manifold
\citep{sorella1998green, sorella2001generalized, carleo2017solving}. Recent sample-space formulations, including minSR, avoid forming the full
\(P\times P\) QGT and instead solve an \(N_s\times N_s\) kernel system, making
SR practical for highly overparameterized NQS
\citep{chen2024empowering, rende2024simple}. SPRING builds on this view with a
Kaczmarz-inspired projected-increment update that reuses information across
minibatches to accelerate convergence \citep{goldshlager2024spring,goldshlager2025sketch}. These
methods improve the cost and convergence of SR, while our focus is the
statistical filtering role of the diagonal shift.

A parallel line of work approximates the QGT or Fisher metric structurally.
KFAC uses Kronecker-factored curvature approximations and has been important in
FermiNet-style electronic-structure VMC
\citep{martens2015optimizing, pfau2020ferminet}. Block-diagonal
QGT methods partition the NQS metric by network layers, preserving intra-layer
curvature while suppressing noisy cross-layer correlations
\citep{shokry2025blockqgt}. These approaches change the metric approximation, but MS-SR keeps the sample-space SR geometry but replaces
the single-shift ridge filter with a data-adaptive mixture of shifted solves.

Our analysis connects SR to random-design ridge regression
\citep{hsu2012random}. In this view, the local-energy target is generally
misspecified by the current tangent space, and the expressivity gap
acts as the irreducible residual noise. Sketch-and-project and Kaczmarz
interpretations clarify the structure of the overparameterized linear solve
\citep{strohmer2009randomized, goldshlager2024spring, goldshlager2025sketch}. The
present work identifies the statistical source of the noise being fitted and
uses that view to design richer spectral filters for SR.

\section{Conclusion}\label{sec:conclusion}

Modern NQS optimization has entered a finite-sample, parameter-rich regime. In
this regime the diagonal shift in SR is not just a safeguard against an
ill-conditioned empirical QGT. It is also the spectral regularizer that
determines how much of the sampled tangent-space target is trusted. By recasting
a single SR step as misspecified ridge regression, we identified the
tangent-space expressivity gap as the effective noise source: it is orthogonal
to the tangent features in population, but it is present in every finite batch
and can be fitted by overparameterized empirical solves.

This viewpoint explains the apparent tension observed throughout the paper.
Larger models can improve SR when gap reduction outweighs additional
finite-sample error, but they can also amplify overfitting when a comparable
gap remains. The exact small-system experiments and the
large fixed-checkpoint SR diagnostics reveal this bias--variance structure
across scales.

The filter perspective also leads to MS-SR, which mixes independent
ridge solves at data-adaptive shifts. The checkpoint ablations separate the
roles of these ingredients. Independent batches provide robust variance reduction, multiple
shifts enlarge the filter class, and stacking adapts the mixture to the local
checkpoint. Paired online continuations and independent endpoint energy
evaluations extend the study to training trajectories, while a separate timing
benchmark quantifies the cost of constructing the updates.

A natural next step is to investigate MS-SR in difficult systems and ask
whether it allows models to reach states that standard optimizers miss. It is
also important to quantify when the extra \(K\)-fold optimizer cost, from solving
\(K\) shifted SR, is worth paying. Such a study requires substantial compute and
system-level engineering, and is a scope of its own.

Overall, overparameterization should not be avoided in NQS, but
it should be treated with the right statistical regularization. Treating SR as a
statistical filtering problem provides both a diagnostic language for
understanding existing optimizers and a design principle for building more
reliable ones.

\begin{ack}
This research was supported by the education and training program of the
Quantum Information Research Support Center, funded through the National
Research Foundation of Korea (NRF) by the Ministry of Science and ICT (MSIT)
of the Korean government (No. RS-2023-NR057243). The author thanks all members of the NQX Lab at \'{E}cole Polytechnique for
their guidance and support in learning about neural quantum states during
the visit. The experimental tools were developed on top of NetKet~\citep{vicentini2022netket}, an open-source neural-quantum-state and variational Monte Carlo library distributed under the Apache License 2.0.
\end{ack}

\bibliographystyle{plainnat}
\bibliography{references}

\appendix
\section{Fixed-Checkpoint SR Regression Identities}\label{app:proofs}

This appendix gives the derivations behind Section~\ref{sec:ridge}. All
statements are checkpoint-local: the wavefunction \(\psi_\theta\), sampling
distribution \(\pi\), QGT, and tangent features are held fixed while only the
Monte Carlo batch used to estimate the SR direction varies. We write
\(\Snorm{u}=u^T S u\). Since \(S\) may be singular, null QGT directions do not
change tangent predictions and have zero \(S\)-seminorm. Throughout,
\(^{\dagger}\) denotes the Moore--Penrose pseudoinverse; Hermitian adjoints are
written explicitly as \(\overline{u}^{\,T}\).

\subsection{Real and complex conventions}

The main text assumes real wavefunctions and real tangent features in the
sampling basis. Then
\[
    S=\Expect_\pi[O_c(x)O_c(x)^T],
    \qquad
    g=\Expect_\pi[O_c(x)H_{\mathrm{loc},c}(x)] .
\]
The fixed-checkpoint population problem is the ordinary least-squares problem
\[
    \min_\delta
    \Expect_\pi\!\left[
        \left(O_c(x)^T\delta-H_{\mathrm{loc},c}(x)\right)^2
    \right],
\]
and its minimum-norm solution is \(\delta^*=S^\dagger g\).

For complex wavefunctions, let
\[
    O_c(x)=\nabla_\theta\log\psi_\theta(x)
    -\Expect_\pi[\nabla_\theta\log\psi_\theta],
    \qquad
    H_{\mathrm{loc},c}(x)=H_{\mathrm{loc}}(x)-\Expect_\pi H_{\mathrm{loc}} .
\]
The Hermitian QGT and force are
\[
    Q=\Expect_\pi[\overline{O_c(x)}\,O_c(x)^T],
    \qquad
    f=\Expect_\pi[\overline{O_c(x)}\,H_{\mathrm{loc},c}(x)] .
\]
For complex parameter increments \(\delta\in\mathbb C^p\), the regression loss is
\(\Expect_\pi|O_c(x)^T\delta-H_{\mathrm{loc},c}(x)|^2\), the normal equations are
\(Q\delta=f\), and the corresponding seminorm is
\(\overline{\delta}^{\,T}Q\delta=\Expect_\pi|O_c(x)^T\delta|^2\). For real parameter
increments in a complex wavefunction, minimizing the same modulus-squared loss
over \(\delta\in\mathbb R^p\) gives $\operatorname{Re}[Q\,\delta]=\operatorname{Re}[f].$
Equivalently, one stacks the real and imaginary residuals and solves an ordinary
real least-squares problem. Thus the real-parameter SR equation uses
\(S=\operatorname{Re}[Q]\) and \(g=\operatorname{Re}[f]\). The derivations below are
written in the real convention. In the complex convention, replace scalar
squares by modulus squares and replace the \(S\)-seminorm by the corresponding
real or Hermitian QGT seminorm.

\subsection{Validation and variance identities}

Let \(\delta^*\) be the population least-squares SR direction and define
\[
    \epsilon(x)=H_{\mathrm{loc},c}(x)-O_c(x)^T\delta^* .
\]
The population normal equations give
\(\Expect_\pi[O_c(x)\epsilon(x)]=0\). Hence, for any fixed direction
\(\delta\),
\[
\begin{aligned}
    \Expect_\pi\!\left[
        \left(O_c(x)^T\delta-H_{\mathrm{loc},c}(x)\right)^2
    \right]
    &=
    \Expect_\pi\!\left[
        \left(O_c(x)^T(\delta-\delta^*)-\epsilon(x)\right)^2
    \right] \\
    &=
    \Snorm{\delta-\delta^*}
    +\sigma_{\mathrm{gap}}^2,
\end{aligned}
\]
where \(\sigma_{\mathrm{gap}}^2=\Expect_\pi[\epsilon(x)^2]\). Thus an
independent held-out residual estimates the \(S\)-excess risk up to the constant
expressivity-gap term.

For a random diagnostic update \(Z=\hat\delta_\lambda\) computed by resampling
an SR batch at the same checkpoint, let \(\mu=\Expect_{\mathcal D}Z\). Adding and
subtracting \(\mu\) gives
\[
\begin{aligned}
    \mathcal E_\lambda
    =
    \Expect_{\mathcal D}\Snorm{Z-\delta^*}
    &=
    \Snorm{\mu-\delta^*}
    +
    \Expect_{\mathcal D}\Snorm{Z-\mu} .
\end{aligned}
\]
The cross term vanishes because \(\Expect_{\mathcal D}[Z-\mu]=0\). This is the
usual squared-bias plus variance decomposition, with distances measured in the
checkpoint QGT seminorm.

Finally, let \(Z_1,\ldots,Z_m\) be independent copies of \(Z\) and
\(\bar Z=m^{-1}\sum_j Z_j\). The standard sample-variance identity holds in the
same seminorm:
\[
    \Expect\left[
        \frac1{m-1}\sum_{j=1}^m\Snorm{Z_j-\bar Z}
    \right]
    =
    \Expect\Snorm{Z-\mu} .
\]
Equivalently,
\[
    \frac1{m-1}\sum_{j=1}^m\Snorm{Z_j-\bar Z}
    =
    \frac1{\binom m2}
    \sum_{a<b}\frac12\Snorm{Z_a-Z_b} .
\]
Replacing each \(S\)-seminorm prediction norm by its empirical held-out average
\( |\mathcal D_{\mathrm{val}}|^{-1}\sum_x (O_c(x)^T\cdot)^2\) gives the
multi-batch diagnostic used in the main text.

\subsection{Spectral filter proxy}

On the non-null QGT subspace, write
\(S=V\operatorname{diag}(s_i)V^T\) with \(s_i>0\), and let
\(\beta_i^*=(V^T\delta^*)_i\). The population ridge direction is
\[
    \delta_\lambda=(S+\lambda I)^{-1}g
    =(S+\lambda I)^{-1}S\delta^* .
\]
Therefore
\[
    \delta_\lambda-\delta^*
    =
    -\lambda(S+\lambda I)^{-1}\delta^* ,
\]
and the exact population shrinkage bias is
\[
    \Snorm{\delta_\lambda-\delta^*}
    =
    \sum_i
    s_i
    \left(\frac{\lambda}{s_i+\lambda}\right)^2
    (\beta_i^*)^2 .
\]

The scalar variance term in \eqref{eq:spectral_risk} is a reference calculation,
not an assumption needed by the diagnostics. To isolate the residual-noise
contribution, write
\[
    \hat g=\hat S\delta^*+\hat\xi,
    \qquad
    \hat\xi
    =
    \frac1{N_s}\sum_{j=1}^{N_s}O_c(x_j)\epsilon(x_j),
    \qquad
    \Expect[\hat\xi]=0 .
\]
If one ignores the covariance-estimation fluctuation in \(\hat S\) and replaces
the empirical inverse by \(A_\lambda=(S+\lambda I)^{-1}\), then
\[
    \hat\delta_\lambda-\delta_\lambda
    \approx
    A_\lambda\hat\xi .
\]
Let
\(\Sigma_\epsilon=\Expect_\pi[O_c(x)O_c(x)^T\epsilon(x)^2]\). The force-noise
proxy is then
\[
    \Expect_{\mathcal D}
    \Snorm{\hat\delta_\lambda-\delta_\lambda}
    \approx
    \frac1{N_s}
    \operatorname{tr}\!\left[
        S A_\lambda \Sigma_\epsilon A_\lambda
    \right] .
\]
Under the scalar, homoscedastic approximation
\(\Sigma_\epsilon\approx\sigma_{\mathrm{gap}}^2 S\), this becomes
\[
\begin{aligned}
    \Expect_{\mathcal D}
    \Snorm{\hat\delta_\lambda-\delta_\lambda}
    &\approx
    \frac{\sigma_{\mathrm{gap}}^2}{N_s}
    \operatorname{tr}\!\left[
        A_\lambda S A_\lambda S
    \right] =
    \frac{\sigma_{\mathrm{gap}}^2}{N_s}
    \sum_i
    \left(\frac{s_i}{s_i+\lambda}\right)^2 .
\end{aligned}
\]

\subsection{Local Fubini--Study metric}

Let \(|\Psi(\theta)\rangle\) denote the normalized variational state and let
\(P_\perp=I-|\Psi(\theta)\rangle\langle\Psi(\theta)|\). The real QGT is the
pullback of the Fubini--Study metric,
\[
    S_{ij}
    =
    \operatorname{Re}
    \left\langle
        P_\perp\partial_i\Psi(\theta),
        P_\perp\partial_j\Psi(\theta)
    \right\rangle .
\]
In the real sampling-basis convention, this equals
\(\Expect_\pi[O_{c,i}(x)O_{c,j}(x)]\). A Taylor expansion of the normalized state
therefore gives, for nearby displacements \(a\) and \(b\),
\[
    1-
    \left|
        \langle \Psi(\theta+a)\mid\Psi(\theta+b)\rangle
    \right|^2
    =
    (a-b)^T S(a-b)
    +
    O\!\left((\|a\|+\|b\|)^3\right).
\]
Taking \(a=-\eta\delta\) and \(b=-\eta\delta'\) gives
\[
    \mathcal I_\theta(\delta,\delta')
    =
    \eta^2
    (\delta-\delta')^T S(\delta-\delta')
    +
    O(\eta^3)
\]
for bounded update directions. Hence the \(S\)-norm excess risk controls the
expected local infidelity between the population SR update and the empirical SR
update, up to the factor \(\eta^2\) and higher-order terms.

\subsection{Sample centering}

Practical SR uses sample-centered log-derivatives and sample-centered local
energies. This is exactly the ridge-regression problem with an unpenalized
intercept,
\[
    \min_{a,\delta}
    \frac1n\sum_{j=1}^n
    \left(
        a+O(x_j)^T\delta-H_{\mathrm{loc}}(x_j)
    \right)^2
    +\lambda\|\delta\|_2^2 .
\]
For fixed \(\delta\), the optimal intercept is
$
    a=\bar H_{\mathrm{loc}}-\bar O^T\delta .
$
Substitution gives
\[
    \frac1n\sum_{j=1}^n
    \left(
        (O(x_j)-\bar O)^T\delta
        -(H_{\mathrm{loc}}(x_j)-\bar H_{\mathrm{loc}})
    \right)^2
    +\lambda\|\delta\|_2^2 .
\]
Thus sample centering is simply the intercept form of the same regression.
In population, the intercept removes only the constant energy/normalization
component. It does not remove the expressivity gap, because
\[
    \epsilon(x)=H_{\mathrm{loc},c}(x)-O_c(x)^T\delta^*
\]
has \(\Expect_\pi[\epsilon(x)]=0\) and
\(\Expect_\pi[O_c(x)\epsilon(x)]=0\). A finite-sample intercept also removes the
sample mean of the residual, replacing \(\epsilon_j\) by
\(\epsilon_j-\bar\epsilon\), but it cannot fit any \(x\)-dependent part of the
gap. These centering terms are lower-order finite-sample fluctuations and do
not change the checkpoint-local bias--variance identities.

\subsection{Correlated Monte Carlo samples}\label{app:sampling}

The population regression identities do not require independent samples within
an SR batch. Autocorrelation does affect the variance of the empirical force.
For a stationary chain, define $\zeta_t=O_c(x_t)\epsilon(x_t)$ and
$\Gamma_\ell=\operatorname{Cov}(\zeta_0,\zeta_\ell)$. If these covariances are
summable, the long-run force-noise covariance is
\[
    \Omega_\epsilon
    =\Gamma_0+\sum_{\ell=1}^{\infty}
      (\Gamma_\ell+\Gamma_\ell^T),
    \qquad
    \operatorname{Cov}(\hat\xi)
    =\frac{\Omega_\epsilon}{N_s}+o(N_s^{-1}).
\]
Under the same fixed-inverse approximation as the spectral proxy, its variance
term therefore becomes
$N_s^{-1}\operatorname{tr}[S A_\lambda\Omega_\epsilon A_\lambda]$.
Replacing $N_s$ by an effective sample size is a useful scalar approximation
when autocorrelation inflates the relevant directions similarly; a single
scalar generally cannot represent the full matrix $\Omega_\epsilon$.
The large-scale diagnostics use MCMC samples and thus already include the
within-batch correlation present in those runs.

Cross-batch dependence is a separate issue. Independent sampler streams are
used for the candidate batches and held-out evaluation. For independent,
identically distributed updates at a fixed shift, uniform averaging reduces
the QGT-norm variance by exactly $K$ even if samples within each batch are
correlated. Correlated candidate batches introduce cross-covariance terms.
Moreover, shifts estimated from one candidate batch and weights learned from
the candidates are data dependent; the exact factor-$K$ statement does not
automatically apply to the complete adaptive MS-SR procedure. Its variance
reduction is assessed by the multi-batch diagnostic.

\section{Experimental Details: Small-Scale Experiments}\label{app:small}

\paragraph{System and bond convention.}
The experiment in Figure~\ref{fig:smallscale_paradox} uses $16$ spin-$1/2$
sites labeled $i=4x+y$, with $x,y\in\{0,1,2,3\}$. Nearest-neighbor bonds are
periodic in both directions. Diagonal bonds join $(x,y)$ to
$(x+1,y\pm1\bmod4)$ only for $x=0,1,2$. There are $32$ nearest-neighbor and
$24$ diagonal bonds, with couplings $J_1=1$ and $J_2=0.5$ multiplying
$\mathbf S_i\cdot\mathbf S_j$, where $\mathbf S_i=\boldsymbol\sigma_i/2$.
The historical bond constructor omitted the eight diagonal bonds across
the $x=3$ to $x=0$ seam. The reported measurements retain that graph rather
than the fully periodic $J_1$-$J_2$ lattice.
No magnetization sector is imposed: all $2^{16}=65{,}536$ states enter the
exact Born distribution and population sums. The archived graph is exposed
explicitly as \texttt{legacy\_nnn\_open\_x} in the code; a separate
\texttt{periodic} option restores the missing bonds for new experiments.

\paragraph{Regression convention.}
The complex wavefunction is retained when computing
$H_{\mathrm{loc}}(x)=(H\psi_\theta)(x)/\psi_\theta(x)$ and the Born
distribution. The offline diagnostic then uses only
$A(x)=\nabla_\theta\operatorname{Re}\log\psi_\theta(x)$ and
$y(x)=\operatorname{Re}H_{\mathrm{loc}}(x)$, with real parameter coordinates.
Writing $A_c$ and $y_c$ for their Born-centered versions, it computes
\[
 S_{\mathrm A}=\Expect_\pi[A_cA_c^T],\qquad
 g_{\mathrm A}=\Expect_\pi[A_cy_c],\qquad
 \delta_{\mathrm A}^*=S_{\mathrm A}^{\dagger}g_{\mathrm A},\qquad
 \sigma_{\mathrm{gap,A}}^2
 =\Expect_\pi[(y_c-A_c^T\delta_{\mathrm A}^*)^2].
\]
For complex states, the full real-parameter QGT additionally contains
$\operatorname{Cov}_\pi(\nabla_\theta\operatorname{Im}\log\psi_\theta)$.
That contribution and the imaginary local-energy target are omitted here.
Thus these measurements are regression excess risks, not full
complex-SR errors or the complete local infidelity in
Eq.~\eqref{eq:fs_local_metric}. The parameter directions may still change
the wavefunction phase; the diagnostic simply does not score that output.
The gap and risk labels in Figure~\ref{fig:smallscale_paradox} refer to this
restricted regression, for which the least-squares decomposition is exact.

\paragraph{Models.}
The underparameterized model is a complex-valued RBM with hidden-to-visible ratio
$\alpha=2$, giving $P=1120$ real parameters and $P/N_s=0.273$ at the diagnostic
sample size $N_s=4096$. The overparameterized model is a complex-valued ViT with
depth $2$, embedding dimension $24$, $4$ attention heads, MLP expansion factor
$4$, and $2\times2$ patches, giving $P=13{,}448$ real parameters and
$P/N_s=3.283$. Parameter counts are real counts, obtained by unrolling each
complex parameter into real and imaginary parts.

\paragraph{Training and checkpoint diagnostics.}
The source checkpoints are produced by VMC with ordinary complex SR, $4096$ Metropolis-local samples per optimization step, learning rate
$10^{-3}$, and training diagonal shift $10^{-3}$. Exact regression summaries are
saved every $10$ optimizer steps. For each diagnostic checkpoint we store
the amplitude gap and target variances, the population-optimal amplitude
direction, and the full-Hilbert-space arrays needed for the offline regression.

\paragraph{Matched-noise protocol.}
For the matched-noise comparison, the RBM and ViT are trained independently and
checkpoint pairs are selected by matching the amplitude gap variance
$\sigma_{\mathrm{gap,A}}^2$. The pair used in Figure~\ref{fig:smallscale_paradox}
is RBM step $490$ and ViT step $10$, with
$\sigma_{\mathrm{gap}}^2=0.8088$ and $0.7957$ respectively. At these fixed
checkpoints we draw $100$ independent training batches of size $N_s=4096$ from
the exact Born distribution and solve the ridge regression at
$\lambda=10^{-9}$.

\paragraph{Matched-state protocol.}
For the matched-state comparison, we first choose a partially converged RBM
checkpoint and then fit a ViT to the same wavefunction by minimizing the exact
Fubini--Study infidelity
\begin{equation}
    \mathcal{I}(\theta_{\mathrm{ViT}})
    =
    1 -
    \frac{
        \left|\sum_x \psi_{\mathrm{RBM}}^*(x)\psi_{\mathrm{ViT}}(x;\theta_{\mathrm{ViT}})\right|^2
    }{
        \left(\sum_x |\psi_{\mathrm{RBM}}(x)|^2\right)
        \left(\sum_x |\psi_{\mathrm{ViT}}(x;\theta_{\mathrm{ViT}})|^2\right)
    } .
\end{equation}
All sums run over the $2^{16}$ basis states, so the objective and diagnostics do
not use stochastic overlap estimates. The pair used in
Figure~\ref{fig:smallscale_paradox} has fidelity $F=0.983626$ between the two
represented complex states. Its exact amplitude gap variances are
$\sigma_{\mathrm{gap}}^2=1.9469$ for the RBM and $0.8161$ for the ViT. We use
$10$ independent Born-resampled training batches of size $4096$ for this
matched-state regression diagnostic.

\paragraph{One-step regression metrics.}
For each resampled training batch \(\mathcal D_b\), the excess-risk
estimate uses the exact amplitude covariance:
\[
    \widehat{\mathcal E}_{\lambda,\mathrm A}
    =
    \frac1B
    \sum_{b=1}^B
    \|\hat\delta_{\mathrm A,\lambda}(\mathcal D_b)
      -\delta_{\mathrm A}^*\|_{S_{\mathrm A}}^2.
\]
We also compute the training-batch residual and the population
real-local-energy regression residual, which adds the constant
amplitude gap variance \(\sigma_{\mathrm{gap,A}}^2\).
Figure~\ref{fig:smallscale_paradox} suppresses the $\mathrm A$ subscripts.

\section{Experimental Details: Large-Scale Experiments}\label{app:large}

This appendix gives the reproducibility details for the two large-scale systems
used in Sections~\ref{sec:large_scale} and~\ref{sec:mssr}: the $L=100$ TFIM
foundation NQS and the $8\times8$ $J_1$-$J_2$ ViT.

\subsection{FNQS on TFIM}

\paragraph{Hamiltonian.}
The periodic one-dimensional transverse-field Ising model is
\[
    H(h)=-h\sum_{i=1}^{L}X_i-\sum_{i=1}^{L}Z_iZ_{i+1},
    \qquad L=100,
\]
with transverse field $h\in[0.8,1.2]$. This interval straddles the critical
point $h=1$.

\paragraph{Foundation NQS objective.}
Following \citet{rende2025foundation}, we train a conditional wavefunction
$\log\psi_\theta(x;h)$ rather than a separate model for each field value. The
field $h$ is supplied as a coupling token to the ViT. Training minimizes the
field-averaged variational energy
\begin{equation}\label{eq:fnqs_loss}
    \mathcal{L}(\theta)
    =
    \Expect_{h\sim p(h)}
    \left[
        \frac{
            \langle\psi_\theta(\cdot;h)|H(h)|\psi_\theta(\cdot;h)\rangle
        }{
            \langle\psi_\theta(\cdot;h)|\psi_\theta(\cdot;h)\rangle
        }
    \right],
\end{equation}
where $p(h)$ is uniform over $R=6000$ training fields in $[0.8,1.2]$.

\paragraph{Model and optimizer.}
The model is a real-valued ViTFNQS with $6$ layers, $d_{\mathrm{model}}=72$,
$12$ attention heads, patch size $b=4$ (effective sequence length $25$), one
coupling input (\texttt{n\_coups}=1), and translational invariance. It has
$P=198{,}144$ trainable parameters. Each SR step uses $N_s=12{,}000$ samples,
organized as $6000$ field replicas with two spin configurations per field. Spin
configurations are drawn with a local Metropolis rule and sweep size $L$. The
optimizer is SGD with fixed learning rate $\eta=0.005$ and minSR/NTK
preconditioning using the \texttt{pinv} linear solver.

\paragraph{Training and checkpoints.}
The source run used for the shared-checkpoint diagnostic protocol is trained
with $\lambda_{\mathrm{train}}=10^{-4}$ for $2000$ steps. Checkpoints are saved
every $100$ steps. The ten diagnostic checkpoints are
$\{1100,1200,\ldots,2000\}$.

\paragraph{Held-out field bank.}
Validation uses a fixed held-out bank of $6000$ transverse fields drawn
uniformly from $[0.8,1.2]$ with seed $0$, disjoint from the training fields.
Each validation pass uses $20$ spin configurations per held-out field, giving
$|\mathcal D_{\mathrm{val}}|=120{,}000$.

\subsection[J1-J2 ViT with symmetry ramp]{$J_1$-$J_2$ ViT with symmetry ramp}

\paragraph{Hamiltonian.}
The two-dimensional system is the $8\times8$ periodic $J_1$-$J_2$ Heisenberg
model
\[
    H
    =
    J_1\sum_{\langle i,j\rangle}\boldsymbol\sigma_i\cdot\boldsymbol\sigma_j
    +
    J_2\sum_{\langle\!\langle i,j\rangle\!\rangle}
    \boldsymbol\sigma_i\cdot\boldsymbol\sigma_j,
    \qquad
    J_1=1,\quad J_2=0.5 .
\]
Here $\boldsymbol\sigma_i$ is the Pauli vector. These raw energy units
are four times those of the same graph with spin operators
$\mathbf S_i=\boldsymbol\sigma_i/2$ and the same numerical couplings.
All energies and energy-squared diagnostics retain these Pauli units.
The graph is fully periodic, with $128$ bonds of each coupling type,
and sampling is restricted to $\sum_i S_i^z=0$.
No Marshall sign rule is imposed.

\paragraph{Model and optimizer.}
The ansatz is the ViT architecture of \citet{rende2024simple}, with depth $6$,
$d_{\mathrm{model}}=48$, $8$ attention heads, MLP expansion factor $4$, and
$2\times2$ patches. The real parameter count is $P=147{,}216$. SR uses the
minSR formulation with Cholesky solves, peak learning rate $\eta=0.01$, a
$100$-step linear warmup, and a cosine decay schedule.

\paragraph{Symmetry ramp.}
Training uses three phases with increasing symmetry:
\begin{itemize}
    \item identity symmetry for $200$ steps with $N_s=8192$,
    \item translations for $3500$ steps with $N_s=8192$,
    \item translations plus $C_4$ rotations for $3500$ steps with $N_s=6000$.
\end{itemize}
The total training length is $7200$ steps. The checkpoint source used for the
MS-SR ablations is the $\lambda_{\mathrm{train}}=10^{-4}$ run. The ten
diagnostic checkpoints are steps
$\{900,1600,2300,3000,3700,4400,5100,5800,6500,7200\}$.

\subsection{Checkpoint-Local Diagnostic Protocol}

\paragraph{Solve-time shift sweep.}
The large-scale diagnostics hold the source checkpoint fixed and vary only the
solve-time shift. For the TFIM U-curve in Figure~\ref{fig:large_scale_ucurve},
the diagnostic grid is
\[
    \lambda\in\{10^{-8},10^{-7},10^{-6},10^{-5},10^{-4},10^{-3},10^{-2},10^{-1}\}.
\]
Each value recomputes SR updates from fresh training batches at the same stored
parameters.

\paragraph{Delta banks.}
For each checkpoint and solve-time shift, we build a bank of independent SR
updates by repeatedly restoring the checkpoint state, drawing a fresh training
batch, computing the SR update, and discarding the parameter update. This makes
the stored directions independent draws from the checkpoint-local law of
$\hat\delta_\lambda$. The TFIM U-curve in Figure~\ref{fig:large_scale_ucurve} uses $m=100$
deltas per shift. The Figure~\ref{fig:mssr_diagnostics} baseline and ablation
comparisons use $20$ resampling repeats per checkpoint.

\paragraph{Validation pass.}
The validation diagnostics are computed by a streamed pass over a fixed
validation block. Local energies and log-derivative Jacobian-vector products are
computed once per validation sample and reused for all stored deltas. The
$J_1$-$J_2$ validation block has $100{,}000$ samples. The TFIM validation block
has $120{,}000$ samples, organized as $6000$ held-out fields times $20$ spin
configurations per field.

\paragraph{Centering and metrics.}
The large-scale implementation uses the energy-gradient convention with
regression target $2H_{\mathrm{loc},c}$ and correspondingly doubled SR
directions. Figures~\ref{fig:large_scale_ucurve}--\ref{fig:mssr_diagnostics}
and the large-scale tables retain these raw residual and variance values:
both diagnostics are four times those under the $H_{\mathrm{loc},c}$
normalization in Section~\ref{sec:ridge}. Relative comparisons, shift selection,
and method ordering are unchanged by this common normalization.
The Pauli-to-spin Hamiltonian conversion for $J_1$-$J_2$ is separate:
at fixed checkpoints, consistently dividing the Hamiltonian and directions
by four divides squared diagnostics by $16$. The figures apply neither
conversion.
For $J_1$-$J_2$, local energies and predictions are centered globally over the
validation block. For TFIM, they are centered separately within each held-out
field, matching the replica structure of the training objective. The validation
residual $\widehat{\mathcal E}_{\mathrm{val}}$ is the mean squared held-out
regression residual for the stored updates. The multi-batch variance
$\mathcal V_{\mathrm{mb}}$ is the centered prediction-space disagreement of the
stored updates around their bank mean; it is computed from
$O_c(x)^T\hat\delta_j$ rather than from raw parameter differences, so flat
parameter directions do not contribute.

\subsection[Historical J1-J2 Online Continuations]{Historical $J_1$-$J_2$ Online Continuations}
\label{app:historical_online}

\paragraph{Source state and paired continuations.}
Figure~\ref{fig:cost_and_online}(a,b) compares five paired continuations from
one shared SR checkpoint at iteration $2500$ of the $8\times8$ $J_1$-$J_2$
model, using the same Pauli Hamiltonian and zero-magnetization sector
specified above. The ViT has depth $8$, embedding dimension $72$, $12$ attention heads,
and MLP expansion factor $4$. It uses only the identity symmetry operation:
the wavefunction is not explicitly symmetrized over lattice translations or
rotations. This is a separate setup from the symmetry-ramp experiment above.
The continuation seeds are $104$--$108$, corresponding in order to pairs
$1$--$5$ in Figure~\ref{fig:cost_and_online}(b). Thus the five pairs sample
continuation variability conditional on one source state, rather than
variability across independently trained source states.

\paragraph{Update construction and budgets.}
Both methods use Cholesky solves, $N_s=8192$ samples per batch, and constant
learning rate $\eta=0.01$. SR uses $\lambda=10^{-4}$. The historical MS-SR
implementation constructs four candidate updates on independent batches at
the fixed shifts $\{10^{-2},10^{-3},10^{-4},10^{-5}\}$, then fits stacking
weights on an additional independent batch. Its callback also executes a
base-driver SR solve on another batch. Each MS-SR update therefore uses five
solves and six sampled batches; $125$ updates use $625$ solves and $750$
batches, compared with $500$ of each for the $500$-update SR control.
This fixed-shift, separate-weight-batch protocol differs from the NTK-quantile,
leave-one-batch-out method in Algorithm~\ref{alg:mssr_appendix}.
The historical runs do not compute NTK quantiles; their additional solve
comes from the base driver. For the NTK-quantile implementation, spectrum
estimation and its reuse within the solver are described separately in
Appendix~\ref{app:cost_protocol}.

\paragraph{Endpoint evaluation and uncertainty.}
Each final state is evaluated independently using two fresh-chain replicas,
with $8{,}388{,}608$ retained samples in total. For pair $i$ we report
$\Delta_i=(E_{\mathrm{MS\text{-}SR},i}-E_{\mathrm{SR},i})/N$, where $N=64$;
energies, differences, and their uncertainties are all in Pauli units.
The individual error bars combine the two methods' Monte Carlo standard
errors in quadrature. The $95\%$ interval for the mean paired difference is
$\overline\Delta\pm t_{0.975,4}s_\Delta/\sqrt{5}$, where $s_\Delta$ is the
sample standard deviation across the five pairs. This interval summarizes
variation across continuations conditional on the shared source state.

\paragraph{Training-curve display.}
The horizontal axis in Figure~\ref{fig:cost_and_online}(a) counts one nominal
SR-equivalent update per SR step and four per MS-SR step; both curves end at
$500$. It excludes the historical callback's extra solve and sampled batches.
The shared point at zero is the mean paired initial observation. Subsequent
training observations are averaged within each continuation in half-open bins
of width $20$ on this nominal axis, excluding the initial observation; the
curves and bands show the mean and $\pm1$ sample standard deviation across
the five continuations. Diamonds at $500$ show the independent endpoint means.

\section{Offline Diagnostics and MS-SR Protocols}\label{app:mssr}

\begin{algorithm}[t]
\caption{Multi-Shift SR (MS-SR): one optimization step}
\label{alg:mssr_appendix}
\begin{algorithmic}[1]
\Require parameters \(\theta_t\); number of shifts \(K\ge2\); batch size \(N_s\);
NTK-spectrum quantiles \(\{q_j\}_{j=1}^K\).
\State Draw independent batches
\(\mathcal D_1,\dots,\mathcal D_K\sim\pi_{\theta_t}\), each of size \(N_s\).
On each batch compute centered features \(O_{c,j}\) and centered local energies
\(H_{\mathrm{loc},c,j}\).
\State Estimate one empirical NTK spectrum
\(\{s_i\}=\operatorname{spec}_{+}(O_{c,1}O_{c,1}^T/N_s)\), and set
\[
    \lambda_j
    \leftarrow
    \operatorname{Quantile}(\{s_i\},q_j),
    \qquad j=1,\dots,K .
\]
\State In parallel, compute
\(\hat\delta_j\leftarrow\minSR(\mathcal D_j,\lambda_j)\) for
\(j=1,\dots,K\).
\State Fit simplex weights $\mathbf w$ by numerically minimizing over
$\mathbf v\in\Delta^{K-1}$ the leave-one-batch-out objective
\[
    \sum_{j=1}^K
    \left\|
        \sum_{i\ne j}\frac{v_i}{1-v_j}O_{c,j}\hat\delta_i
        -
        H_{\mathrm{loc},c,j}
    \right\|^2 .
\]
\State Apply the mixture update:
\[
    \theta_{t+1}
    \leftarrow
    \theta_t-\eta\sum_{j=1}^K w_j\hat\delta_j .
\]
\State \Return \(\theta_{t+1},\boldsymbol{\lambda},\mathbf w\).
\end{algorithmic}
\end{algorithm}

\paragraph{MS-SR shifts.}
All MS-SR ablations in Figure~\ref{fig:mssr_diagnostics} use $K=4$, with
NTK-spectrum quantiles $\{0.9,0.7,0.4,0.1\}$. For these offline diagnostics,
the empirical spectrum is computed on a separately seeded reference batch
and cached once per checkpoint, then reused across all $20$ resampling repeats.
This holds the shift grid fixed while comparing candidate updates and does
not fit the shifts on the final validation data. Algorithm~\ref{alg:mssr_appendix}
shows the online construction, where the spectrum can instead be obtained
from the first candidate batch at the current step, as in the optimized
benchmark of Appendix~\ref{app:cost_protocol}.

Here $O_{c,j}$ has the unnormalized centered features as its rows, so the
factor $1/N_s$ makes the NTK eigenvalues agree with the nonzero eigenvalues of
the empirical QGT. Quantiles use only strictly positive eigenvalues. In the
offline diagnostics, a nonfinite or nonpositive selected quantile falls back
to $\max(s_{\min,+},10^{-12}s_{\max})$; the optimized benchmark additionally
floors every selected shift at this value. For real parameters and complex
features, the corresponding real-stacked feature matrix is used.

\paragraph{Shared-batch and independent-batch constructions.}
The shared-batch variants solve all four shifts on one shared fitting batch. This
preserves the exact finite-sample interpretation as a convex mixture of ridge
filters on a single empirical QGT. The independent-batch variants solve one
shift per independent fitting batch. They no longer correspond to a single
empirical spectral filter, but their sampling errors are decorrelated, giving
the variance-reduction mechanism analyzed in Section~\ref{sec:mssr}.

\paragraph{Stacking weights.}
For MS-SR, weights are selected by leave-one-batch-out stacking. Candidate
update \(\hat\delta_i\) is evaluated on batches that did not produce it, and the
weights are constrained to the simplex. The shared-batch stacked ablation uses the
same candidate deltas as the shared-batch uniform method. TFIM fits the simplex
weights on an independent weight-fit batch. The $J_1$-$J_2$ shared-batch
variants instead split one sampled batch into a fitting part and a held-out
weight-fit part: $6144/2048$ samples through step $3700$, and $4496/1504$
thereafter. Independent-batch methods use the full $8192$ or $6000$ samples
per candidate. Uniform variants set
\(w_i=1/K\). The leave-one-batch-out objective is defined for $w_j<1$:
a simplex vertex would leave no candidate after removing its sole nonzero
weight. The implementation rejects trial weights with
$1-w_j\leq10^{-12}$ rather than evaluating this undefined denominator. The weights are obtained with SLSQP initialized
at the uniform mixture. The objective contains the ratios $w_i/(1-w_j)$ and
is not generally convex; numerical fitting does not guarantee a global
minimum. At $K=2$, the leave-one-batch-out objective is constant for interior
simplex weights, since each held-out prediction uses only one remaining
candidate. The implementation consequently retains the uniform initialization,
explaining the identical uniform and stacked results in that case.

\paragraph{Ablations.}
Figure~\ref{fig:mssr_diagnostics} compares the following checkpoint-local
updates:
\begin{itemize}
    \item \textbf{Standard SR:} one SR solve at $\lambda_{\mathrm{train}}=10^{-4}$.
    \item \textbf{Bagged SR (two variants):} four independent SR solves,
    averaged uniformly at the common shift $\lambda_{\mathrm{train}}$
    or the checkpoint-specific validation-selected $\lambda_{\mathrm{tuned}}$.
    \item \textbf{Shared-batch multi-shift, uniform:} four NTK-quantile shifts on
    one batch, averaged uniformly.
    \item \textbf{Shared-batch multi-shift, stacked:} the same four shared-batch
    candidates with simplex weights fit on held-out samples as described above.
    \item \textbf{Independent-batch multi-shift, uniform:} four independent
    batches, one NTK-quantile shift per batch, averaged uniformly.
    \item \textbf{MS-SR:} four independent batches, NTK-quantile shifts, and
    leave-one-batch-out stacking weights.
\end{itemize}
At each checkpoint, the tuned shift $\lambda_{\mathrm{tuned}}$ minimizes the
mean held-out residual of the four-update bagged average across $20$ resampling
repeats, over $\{10^{-8},10^{-7},\ldots,10^{-1}\}$. The same four sampled
batches are reused across shifts within each repeat. This requires additional
shifted solves and evaluations using the reporting data, making it an oracle
reference. Since the grid includes $\lambda_{\mathrm{train}}$, the selected
mean residual cannot exceed that of training-shift bagging on these data.
The variance is reported at the same selected shift; it is not optimized
separately. Bagging at $\lambda_{\mathrm{train}}$ needs no such search.
For TFIM, $\lambda_{\mathrm{tuned}}=10^{-7}$ at step $1100$ and $10^{-6}$ at
the other nine checkpoints. The $J_1$-$J_2$ choices range from $10^{-4}$ to
$10^{-1}$. We use $\lambda_{\mathrm{single}}$ for the shift minimizing the
single-SR residual on the same eight-point grid. This separate selection is
retained for the single-SR diagnostic in Table~\ref{tab:tfim_baselines} and
the $K$ and source-training-shift robustness experiments below.
The ablation separates the effect of richer spectral filters from the effect of
independent-batch averaging. Bars in the main figure are computed from the same
fixed source checkpoints for each method; error bars use the checkpoint
standard deviation (divisor $10$) divided by $\sqrt{10}$.

\section{Robustness of the Checkpoint-Local Comparisons}\label{app:robustness}

\subsection{Bagging at fixed shifts}\label{app:bagged_shifts}

Table~\ref{tab:tfim_baselines} summarizes the TFIM baselines at the training
shift and at shifts selected for the single-SR or bagged-SR residual.

\begin{table}[htbp]
\centering
\small
\begin{tabular}{lrr}
\toprule
TFIM update rule & $10^3\widehat{\mathcal E}_{\mathrm{val}}$ & $10^3\mathcal V_{\mathrm{mb}}$ \\
\midrule
Single SR, $\lambda_{\mathrm{single}}$ & $9.70\pm2.43$ & $4.71\pm1.24$ \\
Bagged SR, $\lambda_{\mathrm{train}}$ & $9.33\pm2.05$ & $0.61\pm0.13$ \\
Bagged SR, $\lambda_{\mathrm{tuned}}$ & $6.17\pm1.49$ & $1.24\pm0.48$ \\
MS-SR, NTK quantiles & $6.80\pm1.59$ & $1.39\pm0.29$ \\
\bottomrule
\end{tabular}
\caption{TFIM baseline comparison: mean $\pm$ standard deviation across ten
checkpoints (20 resampling repeats per checkpoint). Bagged SR and MS-SR use
four independent batches and four solves per candidate shift.
Here $\lambda_{\mathrm{train}}=10^{-4}$ and $\lambda_{\mathrm{single}}=10^{-6}$
throughout. Bagged-residual tuning selects $10^{-7}$ at step $1100$ and
$10^{-6}$ elsewhere. Both tuned references use the evaluation data for selection;
the additional shift-search cost is excluded from the four-solve count.}
\label{tab:tfim_baselines}
\end{table}

\begin{figure}[t]
    \centering
    \includegraphics[width=\linewidth]{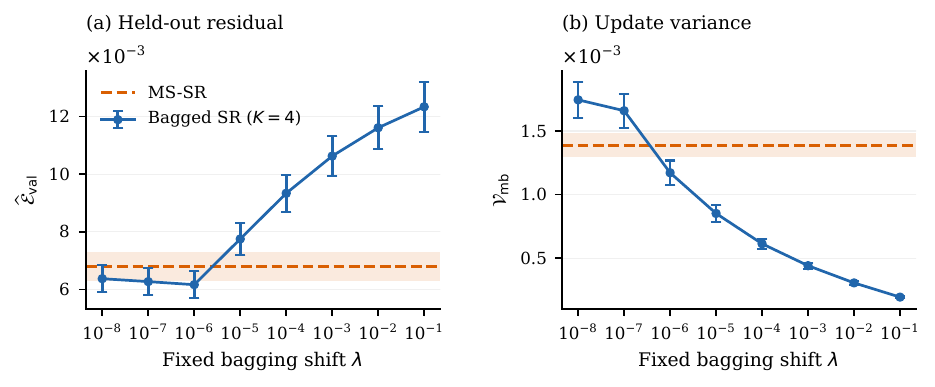}
    \caption{Fixed-shift bagged SR on TFIM ($K=4$). Points are means over ten
    checkpoint-level diagnostics; error bars are standard errors over those
    checkpoints. The dashed line and shaded band show the MS-SR mean and
    standard error on the same checkpoints. The training shift is $10^{-4}$.
    Minimizing the bagged residual selects $10^{-7}$ at step $1100$ and
    $10^{-6}$ at the other nine checkpoints. The complete sweep and
    checkpoint standard deviations are retained in the released summary data.}
    \label{fig:bagged_shift_robustness}
\end{figure}

Figure~\ref{fig:bagged_shift_robustness} sweeps all eight bagging shifts on
the ten TFIM checkpoints, using the same Monte Carlo batches as the $K=4$
ablation. Increasing the shift reduces $\mathcal V_{\mathrm{mb}}$, but the
held-out residual also includes shrinkage bias. MS-SR has mean held-out
residual $0.00680$, compared with $0.00933$ for bagging at the training shift
$10^{-4}$, a reduction of approximately $27\%$. Selecting the bagging shift
per checkpoint gives $0.00617$, which is lower than MS-SR. The fixed shift
$10^{-6}$ is optimal on nine of the ten checkpoints; step $1100$ selects
$10^{-7}$. Thus the comparison depends on the shift available to the
single-shift method; lower variance alone does not imply a better update.

\subsection{Number of candidates and quantile grid}

For this robustness study, $\lambda_{\mathrm{tuned}}=10^{-6}$ denotes a common
reference shift across the ten TFIM checkpoints. It is selected by the
single-SR validation sweep and also minimizes the checkpoint-averaged bagged
residual over the eight tested shifts in the $K=4$ sweep. We retain this
reference as $K$ varies. It also gives the lowest observed mean residual among
$\{10^{-6},10^{-5},10^{-4}\}$ in the saved $K=2,3,5$ comparisons.
Figure~\ref{fig:mssr_diagnostics} instead uses the checkpoint-specific bagged
optimum, which differs from this common reference only at step $1100$.

For a fixed shift and checkpoint, independent uniform averaging leaves bias
unchanged and divides variance by $K$. Hence its predicted residual is
$\widehat{\mathcal E}_{\mathrm{val,single}}-(1-1/K)\mathcal V_{\mathrm{mb,single}}$.
At $\lambda_{\mathrm{tuned}}=10^{-6}$ and $K=4$, this gives
$0.00970-0.75(0.00471)\approx0.00617$, matching the observed bagged residual.
Using the paired single-SR diagnostics for $K=2$, $3$, and $5$ gives agreement
within $0.3\%$. This checks the benefit of averaging at fixed parameters; it
does not compare the resulting update to $K$ sequential optimization steps.

We repeat the TFIM diagnostic with $K\in\{2,3,4,5\}$, using $20$ resampling
repeats per checkpoint and seed-paired comparisons. The quantile grids are
$\{0.9,0.1\}$, $\{0.9,0.5,0.1\}$, $\{0.9,0.7,0.4,0.1\}$, and
$\{0.9,0.7,0.4,0.2,0.1\}$, respectively. Table~\ref{tab:k_robustness} reports
the checkpoint means. The held-out residual improves as $K$ increases for
both MS-SR and bagging; this also increases the number of sampled batches and
SR solves. MS-SR improves on bagging at the training shift at every tested
$K$, while bagging at $\lambda_{\mathrm{tuned}}$ remains stronger on this
TFIM window.

\begin{table}[t]
\centering
\small
\setlength{\tabcolsep}{5pt}
\begin{tabular}{lcccc}
\toprule
Method & $K=2$ & $K=3$ & $K=4$ & $K=5$ \\
\midrule
MS-SR & $8.54/2.28$ & $7.16/1.74$ & $6.80/1.39$ & $6.38/1.27$ \\
Bagged SR, $\lambda_{\mathrm{train}}$ & $9.93/1.22$ & $9.53/0.82$ & $9.33/0.61$ & $9.21/0.49$ \\
Bagged SR, $\lambda_{\mathrm{tuned}}$ & $7.33/2.34$ & $6.56/1.56$ & $6.17/1.17$ & $5.93/0.94$ \\
\bottomrule
\end{tabular}
\caption{Sensitivity to $K$ on TFIM. Each entry is
$\widehat{\mathcal E}_{\mathrm{val}}/\mathcal V_{\mathrm{mb}}$, with both values
in units of $10^{-3}$, averaged over ten checkpoint-level diagnostics.
Here $\lambda_{\mathrm{train}}=10^{-4}$ and $\lambda_{\mathrm{tuned}}=10^{-6}$.
The sampling and solve budget grows linearly with $K$.}
\label{tab:k_robustness}
\end{table}

At $K=4$, we also shift the quantile grid upward or downward
(Table~\ref{tab:quantile_robustness}). The mean held-out residual ranges from
$0.00657$ to $0.00694$, within approximately $3.4\%$ of the default result.
The default quantile grid is fixed across systems; its numerical shifts adapt
to the empirical spectrum, and stacking adapts the mixture weights without a
held-out validation sweep.

\begin{table}[t]
\centering
\small
\begin{tabular}{lcc}
\toprule
Quantile grid & $\widehat{\mathcal E}_{\mathrm{val}}$ & $\mathcal V_{\mathrm{mb}}$ \\
\midrule
$\{0.9,0.7,0.4,0.1\}$ (default) & $6.802\pm1.595$ & $1.389\pm0.291$ \\
$\{0.95,0.8,0.5,0.2\}$ & $6.938\pm1.625$ & $1.469\pm0.283$ \\
$\{0.8,0.6,0.3,0.05\}$ & $6.574\pm1.568$ & $1.369\pm0.321$ \\
\bottomrule
\end{tabular}
\caption{Sensitivity to the $K=4$ NTK quantile grid on TFIM. Values are in units
of $10^{-3}$ and show means $\pm$ standard deviations across ten checkpoint-level
diagnostics, using $20$ resampling repeats per checkpoint.}
\label{tab:quantile_robustness}
\end{table}

\subsection{Source training shift}

To test whether the comparison depends on the source trajectory's diagonal
shift, we repeat the diagnostic at steps $1100$, $1500$, and $2000$ of TFIM
runs trained with $\lambda_{\mathrm{train}}=10^{-3}$ and $10^{-5}$.
The single-SR validation sweep is repeated for each source trajectory and
checkpoint, with $\lambda_{\mathrm{single}}\in\{10^{-5},10^{-6}\}$.
Table~\ref{tab:training_shift_robustness} shows similar mean held-out residuals
for MS-SR and bagging at $\lambda_{\mathrm{single}}$ on both trajectories.
These three-checkpoint comparisons support robustness to the source training
shift within this TFIM setup.

\begin{table}[t]
\centering
\small
\begin{tabular}{clcc}
\toprule
$\lambda_{\mathrm{train}}$ & Method & $\widehat{\mathcal E}_{\mathrm{val}}$ & $\mathcal V_{\mathrm{mb}}$ \\
\midrule
$10^{-3}$ & Single SR, $\lambda_{\mathrm{single}}$ & $36.832\pm6.731$ & $16.348\pm1.470$ \\
 & Bagged SR, $\lambda_{\mathrm{single}}$ & $24.590\pm6.990$ & $4.063\pm0.369$ \\
 & MS-SR & $24.726\pm4.623$ & $5.384\pm0.759$ \\
\midrule
$10^{-5}$ & Single SR, $\lambda_{\mathrm{single}}$ & $2.107\pm0.621$ & $0.700\pm0.179$ \\
 & Bagged SR, $\lambda_{\mathrm{single}}$ & $1.580\pm0.482$ & $0.174\pm0.044$ \\
 & MS-SR & $1.575\pm0.470$ & $0.256\pm0.060$ \\
\bottomrule
\end{tabular}
\caption{TFIM diagnostics on trajectories trained with alternative diagonal
shifts. Values are in units of $10^{-3}$ and show means $\pm$ standard
deviations across the three checkpoint-level diagnostics. Bagged SR and
MS-SR use $K=4$.}
\label{tab:training_shift_robustness}
\end{table}

\section{Compute Resources}\label{app:compute}

\subsection{Steady-state update-cost benchmark}\label{app:cost_protocol}

Figure~\ref{fig:cost_and_online}(c) measures the direct update constructors at the TFIM
step-$2000$ checkpoint, with $L=100$ and $N_s=12{,}000$ samples per candidate
batch. Each run uses one NVIDIA H100 80GB HBM3 GPU on a host with two Intel
Xeon Platinum 8481C processors. All three methods use the same Hermitian
eigendecomposition-based pseudoinverse solver with relative cutoff $10^{-12}$.
SR uses one batch and one solve at $\lambda=10^{-4}$; bagged SR uses four
batches and four solves at that shift, followed by uniform averaging; MS-SR
uses four batches, four NTK-quantile solves, and exact leave-one-batch-out
stacking. The bagged-SR timing corresponds to $\lambda_{\mathrm{train}}$;
it excludes the additional sweep needed to select $\lambda_{\mathrm{tuned}}$.

Timing includes sampling, local energies, candidate solves, and construction
of the final direction. Parameters are held fixed during timing; the common
final parameter write and optional residual diagnostics are excluded.
For MS-SR, the first unshifted NTK eigendecomposition supplies the positive
spectrum quantiles and is reused for the first shifted solve. Stacking reuses
the four candidate batches and their local energies. Its twelve off-batch
candidate predictions are evaluated in four batched Jacobian-vector-product
calls; sufficient statistics are reduced on device before transferring the
small Gram matrices and target projections to the simplex fitter.
The median prediction and fitting times are $0.477$~s and $0.006$~s,
respectively. Reusing the spectrum in this way requires a solver that exposes
its eigendecomposition; an implementation based only on Cholesky or iterative
solves must account separately for spectrum estimation.

We use three paired device/seed configurations, with all methods on the same
physical GPU within each pair. Each method receives two warm-ups followed by
five device-synchronized timed repetitions. Method-independent sampler names
pair the Monte Carlo draws: SR's batch matches the first batch of both
$K=4$ methods, and all four batch fingerprints match between bagged SR and
MS-SR. We take medians over repetitions within each pair, then report the
median and range over the three pair-level summaries. Differences and ratios
are computed on paired repetitions before taking these medians.
The pair-level time ranges are $13.15$--$13.45$~s for SR,
$52.71$--$53.60$~s for bagged SR, and $53.14$--$54.18$~s for MS-SR.
MS-SR adds a paired median of $0.550$~s ($1.027\%$) over bagged SR, with a
three-pair range of $0.437$--$0.579$~s ($0.829$--$1.080\%$).
Thus the methods share the dominant four-batch/four-solve cost, with a small
measured stacking increment in this implementation.

The recorded environment uses Python 3.12.13, JAX 0.7.2, NetKet 3.21.0, and
SciPy 1.18.0. The benchmark implementation is available through
\texttt{sr-filter benchmark}; released timing summaries are under
\texttt{artifacts/paper/figure5/} in the repository.
The recorded device-memory high-water marks are $58.0$~GB for SR and
$58.1$~GB for each $K=4$ method (decimal GB), identical across the three pairs
at this precision. These are process-lifetime/JIT high-water marks and must
not be interpreted as isolated per-update peaks.

\subsection{Historical experiment accounting}

Table~\ref{tab:compute_resources} records the original experiment campaigns.
Its $762$ H100 GPU-hours are a subtotal for those listed campaigns; they do
not include the additional robustness diagnostics, the optimized cost
benchmark, or the final online comparison.

\begin{table}[h]
\centering
\small
\begin{tabular}{p{0.22\linewidth}p{0.22\linewidth}p{0.20\linewidth}p{0.09\linewidth}p{0.09\linewidth}}
\toprule
Experiment & Manuscript use & Worker & Wall time & GPU-hours \\
\midrule
4 by 4 small experiments &
Figures~\ref{fig:smallscale_paradox} and~\ref{fig:mssr_update_tracking} &
\(1\times\) H100 80GB GPU &
\(\approx 30\) h & \(\approx 30\) \\
\midrule
TFIM-FNQS training &
Figure~\ref{fig:large_scale_ucurve} and \ref{fig:mssr_diagnostics} checkpoints &
\(8\times\) H100 80GB GPUs &
\(\approx 1.9\) h & \(\approx 15\) \\
\midrule
TFIM single-shift sweep &
Figure~\ref{fig:large_scale_ucurve} &
\(8\times\) H100 80GB GPUs &
\(\approx 25.5\) h & \(\approx 204\) \\
\midrule
\(J_1\)--\(J_2\) ViT training  &
Figure~\ref{fig:mssr_diagnostics} checkpoints&
\(8\times\) H100 80GB GPUs &
\(\approx 15.1\) h & \(\approx 121\) \\
\midrule
\(J_1\)--\(J_2\) MS-SR ablations &
Figure~\ref{fig:mssr_diagnostics} &
\(8\times\) H100 80GB GPUs &
\(\approx 46\) h & \(\approx 368\) \\
\midrule
TFIM MS-SR ablations &
Figure~\ref{fig:mssr_diagnostics} &
\(8\times\) H100 80GB GPUs &
\(\approx 3.0\) h & \(\approx 24\) \\
\bottomrule
\end{tabular}
\caption{Compute-resource accounting for the original experiment campaigns.
The listed entries sum to approximately \(762\) H100 GPU-hours; this is a
historical subtotal rather than the total for all camera-ready experiments.}
\label{tab:compute_resources}
\end{table}

\clearpage
\section{Online Tracking of a Population SR Reference}\label{app:online_tracking}
\label{par:mssr_online}

\begin{figure}[htbp]
    \centering
    \IfFileExists{figure5_srel_logratio_hamiltonian_boxplots.pdf}{
        \includegraphics[width=0.95\textwidth]{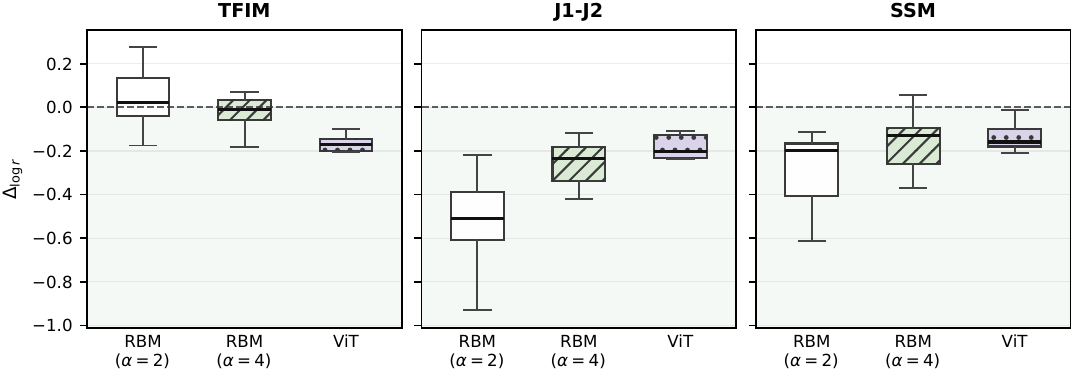}
    }{
        \figureplaceholder{MS-SR trajectory diagnostic placeholder}{
            Boxplots of the paired log-ratio between MS-SR and the stronger of
            SR/SPRING, measured by relative \(S\)-norm error to the exact
            regularized population SR reference over online trajectories.}
    }
    \caption{Online update tracking in exact \(4\times4\) systems. SR, SPRING,
    and MS-SR are trained separately from the same initialization. At each step
    we compare the stochastic update with a full-summation population SR
    reference at shift $10^{-4}$. We plot the full-trajectory log-ratio
    \(\Delta_{\log r}:=\log_{10}(\overline r_{\mathrm{MS}\text{-}\mathrm{SR}}/
    \min\{\overline r_{\mathrm{SR}},\overline r_{\mathrm{SPRING}}\})\). Negative values favor MS-SR. Boxes aggregate random seeds, sample sizes, and
    learning rates for each Hamiltonian--ansatz pair. Iteration counts are
    matched; MS-SR uses four candidate solves plus spectrum and weight-fit
    overhead. This variant smooths its stacking weights across iterations.}
    \label{fig:mssr_update_tracking}
\end{figure}

Figure~\ref{fig:mssr_update_tracking} compares stochastic updates to a
population SR reference along separate online trajectories of SR,
SPRING~\citep{goldshlager2024spring,goldshlager2025sketch}, and MS-SR.
At each optimizer's parameters \(\theta_{a,t}\), exact summation over the
configured Hilbert space gives \(S_{a,t}\) and \(g_{a,t}\), including both
real and imaginary tangent channels for complex wavefunctions. The reference
\(\delta^{\mathrm{ref}}_{a,t}\) is computed with shift $10^{-4}$ and the
smoothed pseudoinverse solver; it is a numerically regularized population
update, rather than the unregularized \(\delta^*\) of Section~\ref{sec:ridge}.
We measure
$
 r_{a,t}=\|\hat\delta_{a,t}-\delta^{\mathrm{ref}}_{a,t}\|_{S_{a,t}}^2
 /\|\delta^{\mathrm{ref}}_{a,t}\|_{S_{a,t}}^2.
$
Eq.~\eqref{eq:fs_local_metric} relates the numerator to local update infidelity
with respect to this reference. The MS-SR variant uses NTK-quantile shifts,
an independent weight-fit batch, and an exponential moving average of its
stacking weights; it differs from the leave-one-batch-out construction in
Section~\ref{sec:mssr}.

The plotted data comprise 96 completed matched configurations across three
\(4\times4\) Hamiltonians (TFIM, \(J_1\)-\(J_2\), and Shastry--Sutherland),
three ans\"atze (RBM \(\alpha=2\), RBM \(\alpha=4\), and a small ViT), and
sample sizes \(N_s\in\{512,1024,2048\}\). The available runs use learning rate
$0.01$ with seeds 0 and 1, and $0.005$ with seeds 0, 1, and 2; the plot retains
only configurations completed by all three methods for \(T=100\) iterations.
For each optimizer we summarize $\overline r_a=T^{-1}\sum_t r_{a,t}$.
Subsection~\ref{app:tracking_protocol} gives the per-panel counts and update variant.

The paired log-ratio in Figure~\ref{fig:mssr_update_tracking} compares MS-SR
with the smaller of the SR and SPRING trajectory-averaged errors. Negative
values favor MS-SR over both baselines. Most completed settings favor MS-SR in
eight of nine Hamiltonian--ansatz pairs; the exception is TFIM with RBM
\(\alpha=2\), where \(4/10\) settings favor it. These are local errors along
distinct trajectories, not distances to one common ideal trajectory or an
equal-budget final-energy comparison.

\paragraph{Hamiltonians and sectors.}
These online experiments use separate constructors from the amplitude
diagnostic in Appendix~\ref{app:small}. All three models have $4\times4$
sites and periodic boundaries, and use Pauli operators.
The $J_1$-$J_2$ Hamiltonian has $32$ nearest-neighbor bonds with $J_1=1$
and all $32$ diagonal bonds with $J_2=0.5$.
The Shastry--Sutherland Hamiltonian has $32$ square-lattice bonds with
coupling $1$ and $8$ orthogonal dimer bonds with coupling $0.8$.
Both Heisenberg models use $\sum_i S_i^z=0$, with
$\binom{16}{8}=12{,}870$ basis states, and no Marshall sign transformation.
The TFIM uses
$H=+\sum_{\langle i,j\rangle}Z_iZ_j-\sum_iX_i$
and the unrestricted $2^{16}=65{,}536$-state space.
Its positive Ising coupling differs from the one-dimensional TFIM convention
in Appendix~\ref{app:large}; on this even bipartite square torus, conjugation
by $X$ on one sublattice reverses the bond sign while leaving the field term
unchanged. The relative QGT errors use the full configured geometry,
rather than the amplitude-only covariance of Figure~\ref{fig:smallscale_paradox}.

\subsection{Protocol and per-panel counts}\label{app:tracking_protocol}

Figure~\ref{fig:mssr_update_tracking} uses the completed matched subset of five
$100$-iteration sweeps: learning rate $0.01$ with seeds $0,1$, and learning
rate $0.005$ with seeds $0,1,2$. The sample sizes are $512$, $1024$, and $2048$.
The plotted asset contains $96$ matched configurations across Hamiltonians,
ansatzes, sample sizes, learning rates, and seeds. A configuration contributes
only when the selected MS-SR variant, SR, and SPRING have complete tracking
records at the target iteration count. Consequently, the panel counts differ
(Table~\ref{tab:tracking_counts}); the figure does not represent a completed
Cartesian sweep of all configurations. These counts and plotted values are
retained from the archived figure-source table. The presently available raw
archive has nine missing method configuration files and four additional
trajectories ending before step $100$; consequently, those entries cannot
currently be reconstructed independently from the available raw trajectories.

The selected variant is \texttt{mssr\_ema\_ntk\_quantile}. It recomputes the four
NTK-quantile shifts at each step, fits simplex weights on an independently
sampled weight-fit batch, and smooths them as
$w_t=0.9w_{t-1}+0.1w_t^{\mathrm{raw}}$ after initialization. These runs use the
research callback with an additional baseline driver solve and spectrum
probe, so the four candidate solves do not represent the complete per-step
cost. This is a comparison at matched iteration counts using this online
variant, with a separate weight-fit batch and smoothed weights; the optimized
exact leave-one-batch-out construction is documented in
Appendix~\ref{app:cost_protocol}.

At every tracked state, the reference direction is computed by exact
summation over the configured Hilbert space with diagonal shift $10^{-4}$ and the
\texttt{pinv\_smooth} solver. It is therefore a population, regularized SR
reference $\delta_{\mathrm{ref}}$, rather than the unregularized
$\delta^*=S^\dagger g$ used in the regression identities. The QGT metric is also
computed by full summation. Each plotted point compares the trajectory-mean
relative QGT-norm error of the selected MS-SR variant with the smaller of the
corresponding SR and SPRING errors. These measurements describe local
direction tracking along the optimizers' own trajectories, rather than a
comparison of final energies at equal sampling or wall-time budgets.

\begin{table}[t]
\centering
\small
\begin{tabular}{lccc}
\toprule
System & RBM, $\alpha=2$ & RBM, $\alpha=4$ & ViT \\
\midrule
TFIM & $4/10$ & $8/11$ & $12/12$ \\
$J_1$-$J_2$ & $11/11$ & $10/10$ & $8/9$ \\
SSM & $12/12$ & $10/11$ & $9/10$ \\
\bottomrule
\end{tabular}
\caption{Per-panel counts for Figure~\ref{fig:mssr_update_tracking}. Each entry
is the number of configurations with lower MS-SR relative QGT-norm error than
both SR and SPRING, divided by the number of completed matched configurations
in that panel. The denominators sum to $96$.}
\label{tab:tracking_counts}
\end{table}

\FloatBarrier

\end{document}